\documentclass[10pt,preprint]{aastex}

\newcommand{\teff}{\ensuremath{T_{\rm eff}}}
\newcommand{\logg}{\ensuremath{log(g)}}
\newcommand{\feh}{\ensuremath{\rm [Fe/H]}}
\newcommand{\kepler}{{\it Kepler} }

\slugcomment{in press at AJ}
\shortauthors{Everett et al.}
\shorttitle{Validation \& Characterization of \kepler Planets}

\begin{document}

\title{High-resolution Multi-band Imaging for Validation and
  Characterization of Small \kepler Planets}

\author{
Mark E. Everett\altaffilmark{1},
Thomas Barclay\altaffilmark{2,3},
David R. Ciardi\altaffilmark{4},
Elliott P. Horch\altaffilmark{5},
Steve B. Howell\altaffilmark{2,6},
Justin R. Crepp\altaffilmark{7},
David R. Silva\altaffilmark{1}
}

\altaffiltext{1}{National Optical Astronomy Observatory, 950 N. Cherry
  Ave, Tucson, AZ 85719, USA}
\altaffiltext{2}{NASA Ames Research Center, Moffett Field, CA 94035,
  USA} 
\altaffiltext{3}{Bay Area Environmental Research Institute, 596 1st Street,
  West Sonoma, CA 95476, USA}
\altaffiltext{4}{NASA Exoplanet Science Institute, 770 South Wilson Ave.,
  Pasadena, CA 91125, USA}
\altaffiltext{5}{Department of Physics, Southern Connecticut State University,
  501 Crescent Street, New Haven, CT 06515, USA} 
\altaffiltext{6}{Visiting Astronomer, Kitt Peak National Observatory,
  National Optical Astronomy Observatory, which is operated by the
  Association of Universities for Research in Astronomy (AURA) under
  cooperative agreement with the National Science Foundation.}
\altaffiltext{7}{Department of Physics, University of Notre Dame,
  225 Nieuwland Science Hall, Notre Dame, IN 46556, USA}

\setcounter{footnote}{7} 

\begin{abstract}
High-resolution ground-based optical speckle and near-infrared
adaptive optics images are taken to search for stars in close angular
proximity to host stars of candidate planets identified by the NASA
\kepler Mission.  Neighboring stars are a potential source of false
positive signals.  These stars also blend into \kepler light curves,
affecting estimated planet properties, and are important for an
understanding of planets in multiple star systems.  Deep images with
high angular resolution help to validate candidate planets by
excluding potential background eclipsing binaries as the source of the
transit signals.  A study of 18 \kepler Object of Interest stars
hosting a total of 28 candidate and validated planets is presented.
Validation levels are determined for 18 planets against the likelihood
of a false positive from a background eclipsing binary.  Most of these
are validated at the 99\% level or higher, including 5 newly-validated
planets in two systems: Kepler-430 and Kepler-431.  The stellar
properties of the candidate host stars are determined by supplementing
existing literature values with new spectroscopic characterizations.
Close neighbors of 7 of these stars are examined using
multi-wavelength photometry to determine their nature and influence on
the candidate planet properties.  Most of the close neighbors appear
to be gravitationally-bound secondaries, while a few are best
explained as closely co-aligned field stars.  Revised planet
properties are derived for each candidate and validated planet,
including cases where the close neighbors are the potential host
stars.
\end{abstract}


\section{INTRODUCTION}\label{sec:introduction}
The NASA \kepler Mission employed a 0.95~m aperture Schmidt telescope
in solar orbit for a total of 4 years (May 2009 -- May 2013).  {\it
  Kepler's} focal plane was filled with 42 CCDs to collect time series
photometry on selected targets in a 115 square degree field.  \kepler
detected transiting exoplanets from a sample of over 150,000 target
stars, most of which fell in the \kepler magnitude range $Kp=8-16$
($Kp$, the \kepler bandpass, spans roughly $430-900$~nm).  The mission
was designed to detect and quantify the population of small planets
orbiting within or near the habitable zones (HZs) of Sun-like stars
\citep{boruckietal10}.  \kepler has produced thousands of candidate
planets, dozens of which are good HZ or near-HZ candidates
\citep{batalhaetal13}.  To help confirm the candidates as true
exoplanets, the mission has relied on ground-based follow-up
observations of the candidate host stars.

The process of producing a list of transiting planets from \kepler
data is a long one.  First, raw pixel fluxes are calibrated
\citep{quintanaetal10}, and light curves are extracted from apertures
and reduced, correcting the flux time series by way of ``cotrending''
to remove variations correlated with ancillary spacecraft data
\citep{twickenetal10}.  At the same time, nearby stars identified in
the \kepler Input Catalog \citep[KIC;][]{brownetal11} are used to
estimate blended (excess) flux in the light curves, and this excess
flux is removed.  Following reduction, the light curves are searched
for any significant periodic events similar to those of transiting
planets \citep{jenkinsetal10b}.  These ``threshold crossing events''
(TCEs) consist of true planet transits and false positives (events
appearing much like planet transits, but attributable to other
phenomena).  False positives include astrophysical sources like
eclipsing binary stars, planets transiting nearby, fainter stars
blended with the KOI star, and instrumental artifacts occurring
(quasi-)periodically in the time series and which coincide when a
light curve is searched on a certain period.  Here and after, ``KOI
star'' refers to the brightest star near the center of the \kepler
aperture as measured in the \kepler bandpass (an unambiguous
definition for the sample in this study).  Cases where the TCEs are
due to planets transiting stars other than the KOI are treated as
false positives because their planetary properties will have been
miscalculated based on adoption of the KOI star properties.  Such
false positives should be removed from the KOI list, if possible, to
maintain it as a well-defined statistical sample.  The TCEs are
subjected to data validation through a series of automated tests
\citep{wuetal10} and human inspection to weed out obvious false
positives.  Those TCEs passing data validation are deemed \kepler
Objects of Interest (KOIs) and given a disposition that identifies
some as planet candidates.  The term KOI can refer to planet
candidates as well as their host stars.  The KOI list forms a large
and relatively clean sample with respect to instrumental false
positives, but still contains a significant number of astrophysical
false positives.  The false positive rate is uncertain, but is likely
to be about $10\%$ \citep{fressinetal13,santerneetal13}.  Follow-up
observations may be used to identify the false positives and more
accurately characterize the host star properties from which planet
properties are derived.  For example, the high resolution follow-up
imaging described here is used to determine the location and
brightness of each star that contributes flux to planet candidate
light curves because {\it Kepler} imaging is optimized only for
photometry (having $4\arcsec$ wide pixels, typical stellar profiles of
$\sim6\arcsec$~FWHM and variably-sized photometric apertures that are
typically several pixels across).

The number of confirmed or validated \kepler planets currently stands
at 965, which is 23\% of the total number of both confirmed and
candidate \kepler planets (4233).  This relatively small fraction
partially reflects the challenges to follow-up observing and analysis
needed to confirm planets with a low level of false positive
probability.

This study presents the analysis of high spatial resolution
observations of 18 KOI stars and the 28 validated and candidate
planets they harbor.  The stellar sample is listed in
Table~\ref{table:spklobservations}.  Each KOI star has been observed
using high resolution optical speckle imaging techniques to search for
or put limits on the brightness of previously-unresolved neighboring
stars.  Many have also been observed in the near-infrared (near-IR)
with adaptive optics imaging with the same goals in mind.  Most of the
host stars have been observed spectroscopically to define their
stellar properties, while the others have stellar properties available
in the literature.  The high-resolution imaging is used to calculate a
validation level for 18 planets around 12 of these stars by
constraining the non-detection of nearby sources.  Two new validated
planetary systems containing 5 planets are designated Kepler-430 and
Kepler-431.  The effects of blending by neighboring stars are examined
and quantified for planets orbiting the 7 affected stars and tests are
performed that help to distinguish whether these neighboring stars are
gravitationally-bound companions or field stars.  These high
resolution imaging and single epoch spectral observations prove to be
an efficient follow-up method for planet validations and refinement of
the planet and host star sample.  Such observations lead to a better
understanding of the sample of small \kepler planets.

\section{CANDIDATE PLANET SAMPLE}\label{sec:sample}

The sample analyzed here is a set of KOI host stars observed with
optical speckle imaging at Gemini North during July 2013.  These
targets were selected from the KOI list at the time on the basis of
two main considerations: (1) they were not previously observed with
high resolution optical imaging at an 8~m or larger telescope and (2)
they hosted a candidate planet having an estimated radius less than
$1.5R_{\bigoplus}$ and/or a predicted planet equilibrium temperature
$T_{eq}<320$~K.  At the time of target selection there was a total of
750 stars hosting at least one planet meeting this size constraint and
20 stars hosting at least one planet meeting the temperature
constraint (temperatures low enough to be considered HZ candidates).
Since that time, planets have been validated for 140 of these 750 host
stars, primarily as part of a validation study of planets in multiple
planet systems \citep{roweetal14}, although most of these are lacking
the high resolution imaging needed to thoroughly investigate their
possible stellar multiplicity.  A total of 25 of the brightest of
these 750 stars was observed (selected to include some with low
equilibrium temperature), but 5 of the stars were subsequently found
by the mission to be false positive events (mostly cases where the
variable was not the KOI, but another star in the aperture).  The
results for two stars of the sample are discussed separately in the
literature: KOI~571 (Kepler-186) by \citet{quintanaetal14} and
KOI~2626 by \citet{ciardietal14}.  The remaining 18 stars discussed
here (Table~\ref{table:spklobservations}) hosted a total of 28
candidates (although some have been subsequently validated).

Along with new observations, analysis of these candidates began by
inspecting ground-based data and \kepler data products available from
the web site of the \kepler Community Follow-up Observing Program
(CFOP)\footnote{https://cfop.ipac.caltech.edu/}.  This included the
$J$-band survey taken at UKIRT (by Phil Lucas) that covers the entire
\kepler field under relatively good seeing conditions
($0.8-0.9\arcsec$~FWHM).  The $J$ images were examined to locate stars
nearby each KOI.  Sources as close as $\sim1\arcsec$ (corresponding to
408~AU at the mean distance of the stellar sample) could be readily
seen in these images, but more importantly they covered areas outside
of the relatively small fields of the follow-up high resolution
images.  Another data product used were the \kepler Mission's data
validation reports that show light curves and statistical tests on
such things as the motion of stellar centroids in and out of transit,
comparison of the depths of odd versus even numbered transits, offsets
of the transit relative to predicted positions for the star, and
in-transit versus out-of-transit pixel flux differences.  The
statistics in the validation reports help determine if any of the
candidates are particularly suspect as false positives
\citep{brysonetal13}.  The candidates discussed hereafter are ``good''
candidates in that the inspection uncovered nothing especially
indicative of false positives.

\section{OBSERVATIONS AND DATA REDUCTION}\label{sec:observations}

\subsection{Speckle Imaging at Gemini North}

Speckle imaging observations were obtained at Gemini North during the
interval UT 25$-$31 July 2013. The Differential Speckle Survey
Instrument (DSSI), a dual-channel speckle imaging system described by
\citet{horchetal09}, was configured with the 692~nm filter (40~nm
FWHM) on the first port and the 880~nm filter (50~nm FWHM) on the
second port. While the mounting of the camera went smoothly, there was
a light-leak problem in the 880~nm channel on the night of 25 July UT,
so the data from that channel was of significantly lower quality and
will not be reported here. The problem was identified and eliminated
by the start of 26 July UT. The pixel scale and orientation were
measured by observing two well-known binary systems, HU~1176
(ie. HIP~83838 or HR~6377) and STT~535 (ie. HIP~104858 or
HR~8123). The known orbital elements from the Sixth Orbit
Catalog\footnote{http://ad.usno.navy.mil/wds/orb6.html} were used to
calculate the position angle and separation at the time of the
observation, and then compared with the raw pixel coordinates, thereby
deriving the scale. Each camera has a slightly different value.  The
final values were determined to be $0.01076\arcsec~{\rm pixel}^{-1}$
for the 692~nm camera and $0.01142\arcsec~{\rm pixel}^{-1}$ for the
880~nm camera. The position angle difference between pixel axes and
celestial coordinates was determined to be $5.69\arcdeg$.

Previous experiences and similar observations were taken during 2012
at Gemini North and are described by \citet{horchetal12}.  Images were
acquired simultaneously in both cameras. The raw data file for each
camera consists of 1000 frames (which is called an ``exposure''); at
least three exposures were taken for each of the objects and were
examined individually and then co-added to achieve the best possible
final result.  While the objects were acquired and centered on the two
detectors with real-time full-frame readout ($512\times512$ pixels),
the science exposures consisted of frames that were
$256\times256$-pixel subarrays, centered on the target.  Each frame
was 60~ms in duration, meaning each exposure represented 1 minute of
integration time.  The choice for the number of exposures taken
generally followed the magnitude of the target, as one would expect,
with the fainter objects receiving more time, but also modified at the
telescope depending on seeing, airmass and other
factors. Table~\ref{table:spklobservations} gives the number of
exposures and the \kepler magnitudes for the systems under study. The
seeing for the run varied between approximately
FWHM$=0.5\arcsec-0.8\arcsec$, with substantial changes from exposure
to exposure for some objects due to weather systems that were in the
area at the time (including Tropical Storm Flossie, which grazed the
Hawaiian Islands on UT 29 and 30 July). Overall, the data from 27 July
represents the bulk of what we present here. This was a relatively
calm night with slightly better seeing than the run as a whole.

\begin{figure}
\epsscale{0.85}
\plotone{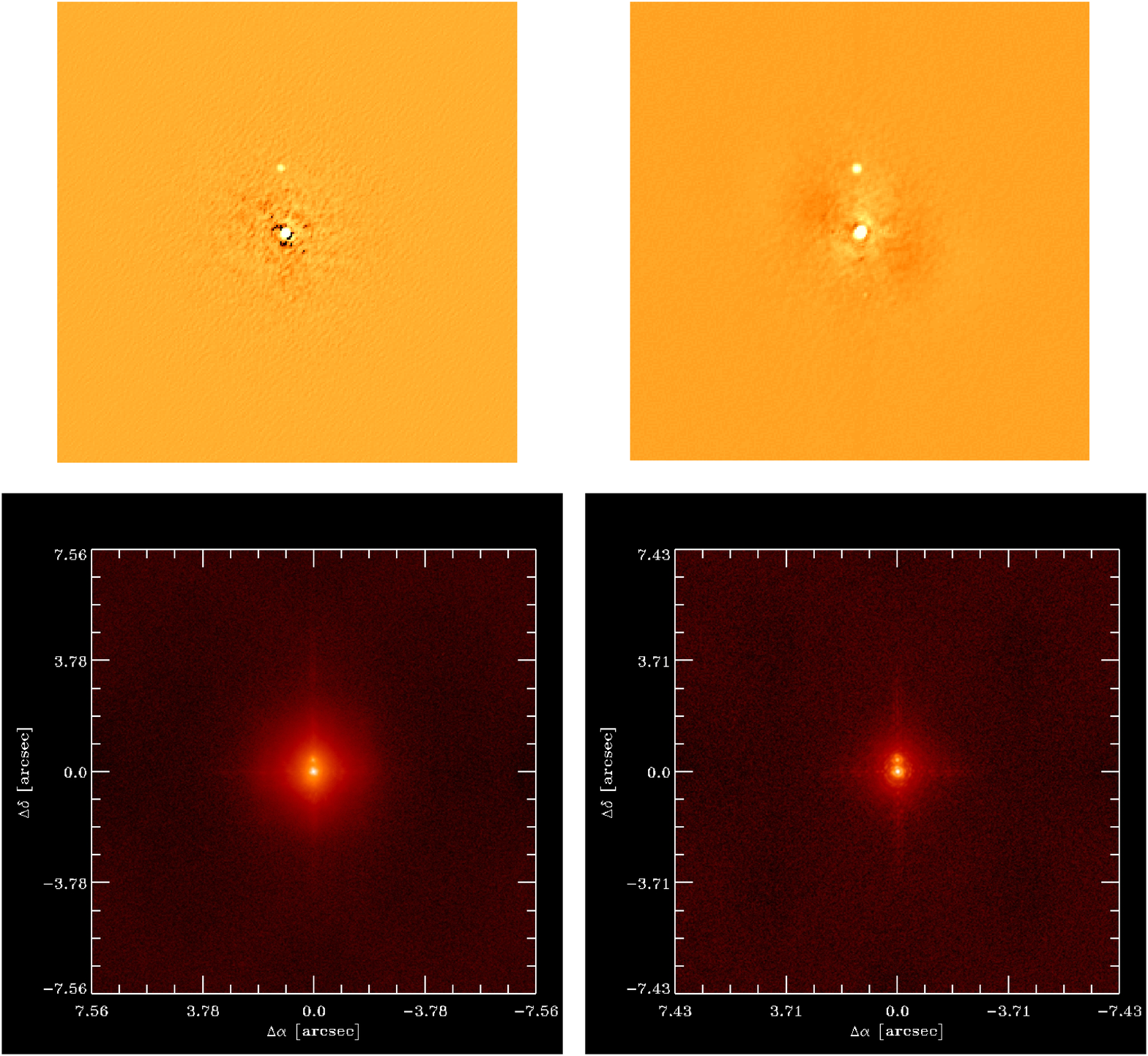}
\caption{
Example high resolution imagery of KOI~1964 and its surroundings in 4
filters.  The upper two panels are reconstructed images from speckle
observations at 692~nm (upper left) and 880~nm (upper right) taken at
Gemini North.  The lower two panels are adaptive optics images at $J$
(lower left) and $Ks$ (lower right) taken at the Palomar Hale
Telescope.  Each image is oriented with North at the top and East to
the left.  The speckle images are $1.8\arcsec\times1.8\arcsec$ and the
adaptive optics images are approximately $15\arcsec\times15\arcsec$ as
seen by the scales.  A faint neighbor star is detected $0.4\arcsec$ to
the north of the brighter KOI star.
} \label{Fig:KOI1964_4filters}
\end{figure}

The basic methodology for speckle data reduction has been described in
previous papers, e.g. \citet{howelletal11} and \citet{horchetal12}.
The latter deals specifically with Gemini data taken in 2012.  It
is based on Fourier analysis of correlation functions made from the
raw speckle data frames. The autocorrelation is used to estimate the
modulus of the object's Fourier transform. A point source observation
is required to deconvolve the point spread function (which amounts to
a division in the Fourier domain). The triple correlation function can
be used to generate the phase of the object in the Fourier
plane. Combining these two functions, an estimate of the Fourier
transform of the object is obtained.  This is then low-pass filtered
with a Gaussian function and inverse transformed to arrive at the
final reconstructed image with a diffraction-limited resolution of
FWHM$\simeq0.02\arcsec$.  Example reconstructed speckle images
centered on the double source KOI~1964 are shown in
Fig.~\ref{Fig:KOI1964_4filters}.

For \kepler follow-up observations, we use the reconstructed images to
measure the limiting magnitude difference of each observation as a
function of distance from the primary star, that is, it is an estimate
of the brightest star that could be missed as a function of separation
from the primary. As shown in the previous papers, these curves are
generally monotonically increasing as a function of separation,
meaning that the limiting magnitude near the central star is lower
than farther away from the star. Up to the present, we have published
$5\sigma$ confidence limits as a function of separation, using all
peaks in the reconstructed image to generate a mean and standard
deviation of the mean of the peak values. A detectable companion star
then must have a peak value larger than the mean plus 5$\sigma$. For
Gemini data, the results on fainter targets from our run in July 2012
generally showed two image artifacts that were undesirable in the
final reconstructed image: a faint cross pattern centered on the
target, and correlated noise patterns over length scales of
$\sim0.05-0.10\arcsec$. These effects can combine to give a
detectability curve with non-Gaussian distribution of peak heights
and/or a non-monotonic nature as a function of separation.

We have studied these two effects in the Fourier plane and developed
two strategies to reduce their appearance in reconstructed
images. First, the cross pattern on the image plane maps to a cross on
the Fourier plane, which can be cleanly seen in a region beyond the
diffraction limit and removed by replacing the pixel values in the
cross with an average of pixel values on either side. Second, the
correlated noise appears to be reduced when the point source used to
do the deconvolution step is a better match to the point spread
function of the target star. Therefore, we have developed an algorithm
to ``fine-tune'' the shape of our point source observation in the
Fourier plane based on estimating the difference in dispersion
expected for the point source observation and the science target
(which is a function of observation time and sky position), and
calibrating out the point source dispersion accordingly. These
techniques appear to yield reconstructed images which are free of the
cross and whose noise peaks have a more Gaussian distribution.

\subsection{Near-IR AO Imaging}

Ten of the KOIs were observed with near-IR adaptive optics (AO) in the
$J$, $K^\prime$ and $Ks$ filters either at the Lick Observatory Shane
3.5~m, the Palomar Observatory Hale 5~m or the 10m Keck-II Telescope
(see Table~\ref{table:AOobservations}), as part of a general infrared
AO survey of KOIs \citep[e.g.,][]{roweetal14,marcyetal14,adamsetal12}.

Targets observed with the Lick, Palomar, or Keck AO systems utilized
the IRCAL \citep{lloydetal00}, PHARO \citep{haywardetal01}, or NIRC2
\citep{wizinowichetal04,johanssonetal08} instruments respectively. The
observations were made in the $J$ filter for the Lick observations,
the $J$ and $Ks$ filters for the Lick and Palomar observations, and
the $K^\prime$ filter for the Keck observations.

The targets themselves served as natural guide stars and the
observations were obtained in a 5-point quincunx dither pattern at
Lick and Palomar, and a 3-point dither pattern at Keck to avoid the
lower left quadrant of the NIRC2 array.  Five images were collected
per dither pattern position, each shifted $1\arcsec$ from the previous
dither pattern position to enable the use of the source frames for
creating the sky image.  The IRCAL array is $256\times256$ with
75~mas pixels and a field of view of $19.2\arcsec\times19.2\arcsec$,
the PHARO array is $1024\times1014$ with 25~mas pixels and a field of
view of $25.6\arcsec\times25.6\arcsec$, and the NIRC2 array is
$1024\times1024$ with 10~mas pixels and a field of view of
$10.1\arcsec\times10.1\arcsec$.

Each frame was dark subtracted and flat fielded and the sky frames
were constructed for each target from the target frames themselves by
median filtering and coadding the 15 or 25 dithered frames. Individual
exposure times varied depending on the brightness of the target but
typically were $10-30$~seconds per frame.  Data reduction was
performed with a custom set of IDL routines.

Aperture photometry was used to obtain the relative magnitudes of
stars for those fields with multiple sources.  Point source detection
limits were estimated in a series of concentric annuli drawn around
the star.  The separation and widths of the annuli were set to the
FWHM of the primary target point spread function.  The standard
deviation of the background counts is calculated for each annulus, and
the $5\sigma$ limits are determined within annular rings \citep[see
  also][]{adamsetal12}.  The PSF widths for the Lick, Palomar, and
Keck images were typically found to be 4~pixels for the three
instruments corresponding to $0.3\arcsec$, $0.1\arcsec$, and
$0.04\arcsec$~FWHM respectively.  Typical contrast levels are
$2-3$~magnitudes at a separation of 1~FWHM and $7-8$~magnitudes at
$>5$~FWHM with potentially deeper limits past 10~FHWM.  An example of
AO imaging done at Palomar toward KOI~1964 is shown in
Figure~\ref{Fig:KOI1964_4filters}.

This study includes observations in both $K^\prime$ and $Ks$ filters.
The $K^\prime$ filter differs only slightly from $Ks$ (with central
wavelengths of $2.12~{\mu{\rm m}}$ and $2.15~{\mu{\rm m}}$
respectively).  Because of this, the differential magnitudes of stars
measured in either filter are treated as equivalent since any
differences are expected to be slight.  For calculations and
modeling, the $Ks$ bandpass is used.

\subsection{Spectroscopy at NOAO Mayall 4m}\label{subsec:spectroscopy}

Most of the KOI host stars (16 of 18) were observed spectroscopically
at the National Optical Astronomy Observatory (NOAO) Mayall 4m
telescope at Kitt Peak during the 2010 and 2013 observing seasons.
Table~\ref{table:spectraobs} lists the 11 spectra actually used to
determine stellar properties (other stars were too cool, too hot, or
have published asteroseismology measurements of stellar properties as
discussed in \S\ref{subsec:literatureproperties}).  The stars were
observed with integration times of $5-15$ minutes using the long-slit
spectrograph RCSpec setup to disperse the spectra with
0.072~nm~pixel$^{-1}$ at a nominal resolution of $\delta\lambda =
0.17$~nm.  The wavelength coverage with the best calibrated fluxes was
approximately $380-490$~nm.  More details of this observing program
are discussed in \citet{everettetal13}.

Spectral frames are reduced in the manner described by
\citet{everettetal13}.  Briefly, the overscan bias is subtracted and
trimmed off each frame.  Bias frames and flat field frames are then
combined, with outlier rejection, to form a master residual bias image
and flat.  These master frames are applied to each observation in the
usual manner.  Stellar spectra are extracted using an aperture that
traces the stellar image across the CCD and sky-subtracted using night
sky spectra extracted from areas of the slit containing sky.
Wavelength calibration is provided by an arc lamp exposure at each
pointing and flux calibration is done using an observation of a
spectrophotometric standard star along with a Kitt Peak extinction
curve scaled to the airmass of each observation.  Since focus changes
significantly across the CCD, only the best focused portion of the
spectrum is used for analysis ($\lambda=460-489$~nm where the focus is
tight and important spectral features like H$\beta$ are found).

\section{PROPERTIES OF THE KOI STARS}\label{sec:stellarproperties}

The properties of the candidate host stars are estimated in a number
of ways.  For most stars, a newly acquired spectrum, taken at the
Mayall 4m telescope, is available as discussed in
\S\ref{subsec:spectroscopy}.  In other cases, values are obtained from
the literature and are variously based on asteroseismology,
photometry, or spectral analysis used in conjunction with light curve
fits.  Of all stellar properties, the radius is the most fundamental
for characterizing transiting exoplanets because it is used to derive
the planet radius.

It is worth noting that a number of the candidate host stars have
neighboring stars closeby.  When the apparent separations are small
enough, the neighbors can affect both the follow-up photometry and
spectroscopy as flux from the neighbor is introduced into the data.
However, in most cases the neighbors are at least several magnitudes
fainter and so the contamination is slight.  To determine the
properties of both the KOI star and its neighbors, we take a two-step
approach: First, the properties of the KOI star are established from
asteroseismology, if available, otherwise spectroscopy or, lastly,
photometry when that is the only available source.  Second, once the
properties of the KOI star are established, the properties of the
neighbors are estimated photometrically as will be discussed in
\S~\ref{subsec:neighborproperties}.  In most cases, the photometry of
the neighbors is measured relative to the KOI star, so determining the
properties of the neighbors depends on first characterizing the KOI
star.

\subsection{New Spectroscopic Properties}

In the case of the KOIs observed spectroscopically at the Mayall 4m
telescope, an estimate for \teff, \logg\ and \feh\ is made in the
manner described in detail by \citet{everettetal13}.  Very briefly,
each spectrum is iteratively fit to a grid of synthetic model spectra
taken from \citet{coelhoetal05}, who parameterized their models using
these three properties.  The spectral models of \citet{coelhoetal05}
are based on the stellar atmosphere models of
\citet{castellikurucz03}, and were chosen by \citet{everettetal13}
from among the publicly available model spectra for their
well-sampled grid in parameter values.  The model fitting method is
calibrated using a set of similar spectra taken of test stars whose
properties were well known {\it a priori}.  Parameter uncertainties
for this method are based on the degree to which the fitted properties
of the test star set matched their {\it a priori} values.  The \teff,
\logg\ and \feh\ values from these spectra are listed in
Table~\ref{table:stellarproperties} and marked as coming from
Reference 1 or 3.  A mass, radius and luminosity is determined later
for these stars based on isochrone fits (see
\S\ref{subsec:isochronefits}).

\subsection{Properties from the Literature}\label{subsec:literatureproperties}

For some KOIs, we have no 4m spectrum or the star was such that it
could not be fit (these spectral fits were reliable only within the
effective temperature range $4750K<T_{\rm eff}<7200K$).  For these stars,
values of \teff, \logg, and \feh\ are taken from the literature.  The
values adopted (in Table~\ref{table:stellarproperties}) are those
listed in the stellar properties catalog of \citet{huberetal14} which
contains ``best available'' properties for almost all of the stars
targeted by \kepler.  It includes properties of very well
characterized stars alongside those based on photometry alone
(generally the least reliable method of characterization).
For those stars with only photometry, like the hot star
KOI~3204, \citet{huberetal14} derive new stellar properties, first by
identifying any giants using asteroseismology, then finding
\teff\ from the available photometry.  They determine other parameters
with a Bayesian statistical analysis that includes
empirically-motivated priors on \feh\ and \logg\ that help constrain
photometric fits of model spectra to optical and near-infrared colors.
For other stars, \citet{huberetal14} rely on existing data as inputs
to the Bayesian analysis.  The properties of the cool stars KOI~3255
and KOI~3284 are calculated based on photometrically-derived
properties from \citet{pinsonneaultetal12} and
\citet{dressingcharbonneau13} respectively.  For KOI~1964, the
constraints are provided by \citet{batalhaetal13} and are based on the
light curve and spectroscopic fitting techniques described by
\citet{buchhaveetal12}.  Three of the KOI stars (KOIs~268, 274, and
1537) have been analyzed both asteroseismologically and
spectroscopically by \citet{huberetal13} who provide quite accurate
and precise values for \teff, \feh, \logg, $R_\star$ and $M_\star$.
For these stars, the mass and radius are the literature values.  For
all other stars the radii and masses are determined
from new isochrone fits described next.

\subsection{Properties from Isochrone Fits}\label{subsec:isochronefits}

A new isochrone fitting procedure has been developed for this study to
determine the stellar properties for both KOI stars and any
potentially bound secondaries (see \S\ref{subsec:neighborsbound} for a
discussion of neighboring stars' properties).  For the purpose of
isochrone fitting, each KOI star is described by the set of most
probable values for the same three properties (\teff, \logg, \feh)
with a probability distribution described by half of a normal
distribution each for the positive and negative uncertainties (which
may differ).

A set of Dartmouth isochrones \citep{girardietal05} is
constructed using the interpolation software provided with
their distribution.  The isochrones span an age range between
$1-13$~Gyr at 0.5~Gyr intervals and metallicity (\feh) range between
$-0.4$ and $+0.4$ with steps of 0.02~dex with no $\alpha$-element
enhancements.  To obtain a finely-sampled set of stellar mass points
defining each isochrone (where the original isochrones had some large
gaps), new points are created using linear interpolation such that the
final intervals between successive stellar masses never exceeds
0.02$M_\odot$.

To find the properties of the primary star, a probability level is
assigned to each mass point in the set of isochrones based on its
location in the (\teff, \logg, \feh) probability distribution.  The
mass point with highest probability and the extent of the parameter
space mapped out by those points whose probabilities fall inside a
certain threshold level define the central values and $1\sigma$
uncertainties in the other stellar properties (e.g., $R_\star$ and
$M_\star$ as listed in Table~\ref{table:stellarproperties} and
absolute magnitudes as discussed later).  Exceptions to this (for
Table~\ref{table:stellarproperties}) were made for stars with
asteroseismology, whose masses and radii are supplied in the available
literature.

The precision to which stellar radius is estimated varies between the
different techniques.  Table~\ref{table:stellarproperties} lists 12
KOIs with stellar radii derived from spectra without the input of
asteroseismology.  The mean uncertainty in stellar radius for these
stars is 16.9\% (averaging all plus and minus uncertainties together).
There are three stars with stellar radii based solely on photometric
colors with a mean uncertainty of 22.8\% (with uncertainties varying
greatly among the sample).  The three stars with properties based on
asteroseismology have a much lower mean radius uncertainty of 2.3\%,
illustrating the impact of this technique.

\subsection{Magnitudes in the 692nm and 880nm Filters}\label{subsec:dssimags}

\begin{figure}
\epsscale{0.67}
\plotone{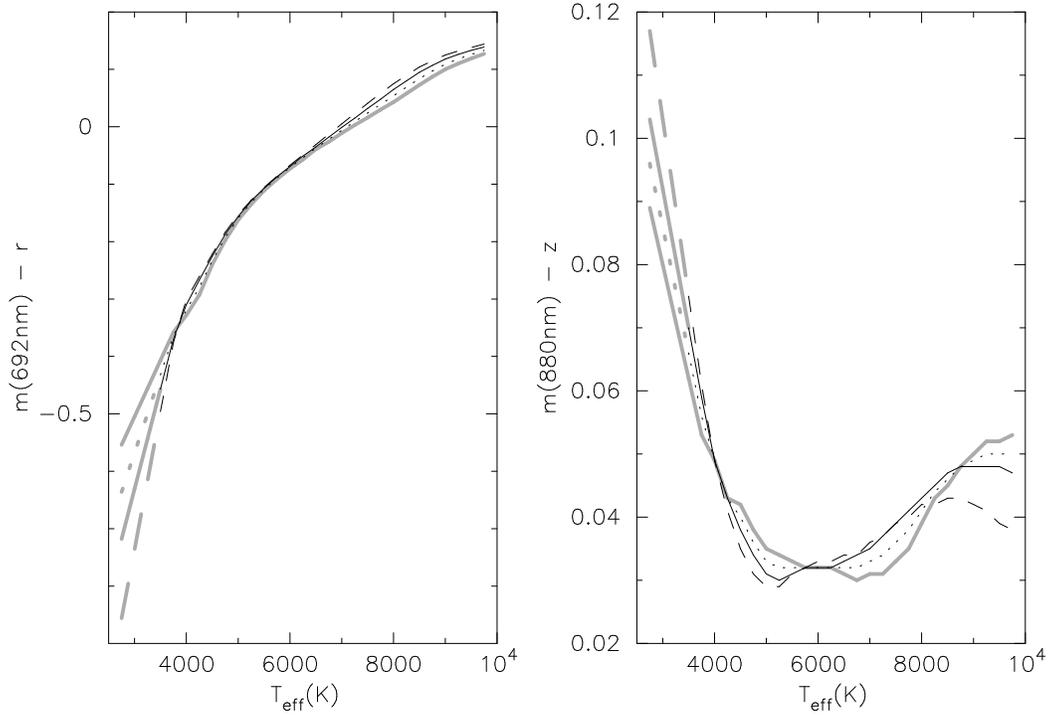}
\caption{
Relationships between stellar effective temperatures and colors
relating the 692~nm to the SDSS~$r$ filter (left panel) and the 880~nm
to the SDSS~$z$ filter (right panel).  These colors are calculated
based on model stellar spectra for ${\rm 3500\leq T_{eff}\leq10^4}$~K
at \logg\ values of 4.0 (black dashed line), 4.5 (black solid line),
and 5.0 (black dotted line).  Linear extrapolations from these curves
are used to predict the colors for ${\rm T_{eff}<3500}$~K and a curve
at ${\rm log(g)=5.5}$ that are not covered by the model spectra.
Solar metallicity, ${\rm [Fe/H]=0}$ is used here.  The extrapolated
curves are shown in grey.
} \label{Fig:color_vs_Teff}
\end{figure}

The Dartmouth isochrones already predict absolute $Kp$, $B$, $V$,
SDSS~$griz$, $J$, and $Ks$ magnitudes, but not magnitudes for the
specialized 692nm and 880nm filters used in speckle imaging.  To add
absolute magnitudes for the 692nm and 880nm filters to the isochrone
data, color$-$\teff\ relationships are derived that relate these
magnitudes to SDSS magnitudes.  These color$-$\teff\ relationships are
calculated based on solar metallicity model spectra published by
\citet{munarietal05}, the filter transmission curves, the QE curve of
the DSSI CCDs, an atmospheric extinction curve for Mauna Kea at the
typical observing airmass of 1.3, and the AB magnitude system.  The
color$-$\teff\ relationships between the SDSS magnitudes and speckle
imaging filters are shown in Figure~\ref{Fig:color_vs_Teff}.

Because the lowest \teff\ in the model spectra of \citet{munarietal05}
was 3500~K, the color$-$\teff\ relationship is linearly-extrapolated
down to 2750~K (although the lowest \teff\ actually found in the
isochrones is $\sim3000$~K).  Additionally, to obtain magnitudes for
stars with $log(g)>5$, a $log(g)=5.5$ curve is found by linear
extrapolation of the colors predicted at $log(g)=4.5$ and 5.0.  These
extrapolations are indicated in Figure~\ref{Fig:color_vs_Teff} with
light grey lines.  For any star defined by \logg\ and $T_{\rm eff}$,
the absolute magnitudes in the speckle bandpasses can now be found by
interpolating between the two bounding curves in the
color$-$\teff\ relationships given their absolute $g$ or $z$
magnitudes.  These calculations are done assuming solar metallicity
models for each star.  Metallicity has a noticeable, but small effect
on colors for stars cooler than 4000~K which increases with decreasing
effective temperature.  At \teff$=3000$~K, there are color differences
of $\sim0.02$ in $880 - z$ and $\sim0.15$ in $692 - r$ when comparing
\feh$=-0.5$ models to solar metallicity models.  Thus, there is
additional uncertainty in modeling fluxes in the speckle filters
among cool stars.  This mainly impacts a few of the faint neighbor
stars discussed in \S\ref{sec:companions}.

\section{FALSE POSITIVE PROBABILITY ANALYSIS}\label{sec:fpp}

A false positive probability is determined for planet candidates
orbiting each KOI star that is sufficiently isolated from detectable
neighbors such that the KOI star is the only {\it detected} star near
the source of the candidate signal.  This analysis compares the
probability of the KOI being a planet host to the probability that a
nearly co-aligned and fainter field star is the source of a false
positive signal (ie. an eclipsing binary or transiting planet).  The
scenario of an unresolved triple KOI in which two components form an
eclipsing pair is not considered here because for these it is
difficult to calculate some of the complex scenarios for a given
system.  Scenarios such as these have been considered by
\citet{fressinetal13} who found that the incidence of false positives
attributable to an eclipsing secondary component in a hierarchical
triple stellar system are quite low, especially among candidates of
Neptune and smaller planets like in the sample considered here.

We estimate the false positive probability for each planet candidate
by integrating the parameter space not excluded by \kepler data or
follow-up observations with respect to a Galactic model. This method
is based on the approach described in \citet{barclayetal13} and
\citet{wangetal13,wangetal14}.  The information used to restrict the
parameter space is the transit depth, the \kepler out-of-transit pixel
response function (PRF) centroid statistic and the 692~nm Gemini
speckle and any near-IR AO observations of the star.  The PRF is the
observed appearance of point sources and depends on the PSF produced
by the \kepler optics, spacecraft jitter, focus, and spectral class of
a given point source (although the latter effect is not considered
significant enough to treat individually).  The measured PRF and its
centroid statistic \citep{brysonetal13}, the quarter-by-quarter
standard deviation between a stellar centroid in an out-of-transit
image and the difference image between in-transit and out-of-transit
light curve points are products of the \kepler pipeline
\citep{tenenbaumetal13,tenenbaumetal14}.

The transit depth provides a limit on the faintest star that could
produce a false positive signal matching the light curve. This comes
from assuming a total eclipse by a background eclipsing binary star of
identical components, which would produce a 50\% eclipse depth.  This
maximum eclipse depth is adopted under the expectation that for
more general binaries with unequal mass components, the larger star
will be brighter in the \kepler bandpass.  For a maximum eclipse depth,
the background star, outside of eclipse, would be ${\Delta}Kp_{max}$
magnitudes fainter than the KOI and the observed transit depth can be
expressed in terms of $\delta$, the KOI's fractional transit depth:
${\Delta}Kp_{max}=-2.5{\times}log_{10}(2\delta)$.  For example, if the
observed transit depth were 100~ppm, this could be induced by a
eclipsing binary of at most $Kp=9.25$~magnitudes fainter than the KOI.
Our estimates of the transit depth are taken from the \kepler data
analysis pipeline
\citep{jenkinsetal10a,tenenbaumetal13,tenenbaumetal14}.  Values for
${\Delta}Kp_{max}$ are listed in Table~\ref{table:validations}.

The \kepler out-of-transit PRF centroid statistic is used to set an
exclusion radius for each planet candidate.  Any star outside this
exclusion radius is excluded from being the source of the candidate
transit signal because any such source outside this radius would
produce a larger centroid statistic.  To find the exclusion radius, we
use a $3\sigma$ threshold where $\sigma$ is the PRF statistic
discussed above.  These exclusion radii establish which KOI stars are
sufficiently isolated for the analysis as well as restrict the area
inside of which false positives are modeled.  Values for the exclusion
radius, $r_{ex}$, are listed in Table~\ref{table:validations}.  In the
case of KOI~2311, no exclusion radius could be determined due to the
lack of PRF centroid data on this star.  As discussed in
\S\ref{sec:companions}, there are 8 candidate planets transiting 5 KOI
stars that show a neighboring star within the exclusion radius.  These
candidates, plus the two of KOI~2311 are excluded from the validation
calculation.

\begin{figure}
\epsscale{0.67}
\plotone{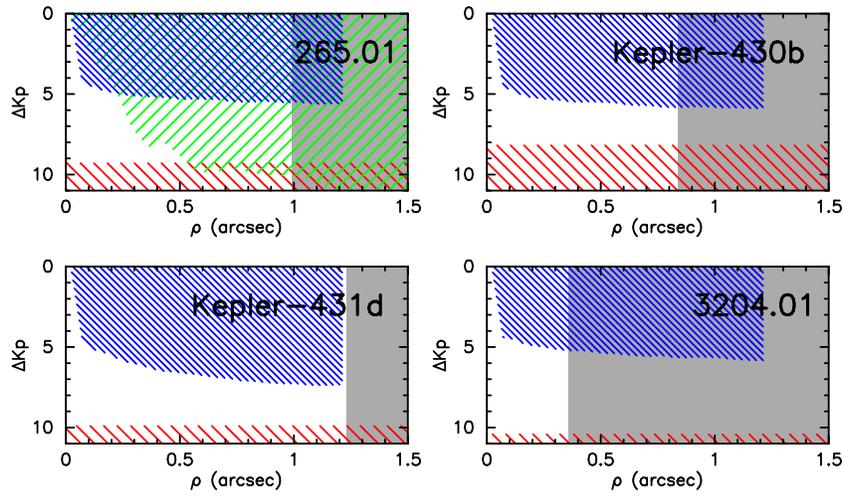}
\caption{
Areas of parameter space where data excludes background stars as the
source of a false positive planet signal are shown in terms of the
relative \kepler magnitudes fainter (${\Delta}Kp$) and angular
separation ($\rho$) with respect to 4 example host stars.  Stars are
excluded from the blue hashed areas by 692~nm speckle imaging and the
green hashed area by $Ks$-band adaptive optics imaging.  The red
hashed area shows regions below the minimum brightness star
(${\Delta}Kp_{max}$) that could produce the observed transit depth.
The grey area blocks out areas beyond the exclusion radius set by
\kepler PRF centroids statistics.  Predictions for the number of
background stars that would lie in the remaining white area are used
to validate the planets.  Labels in each panel indicate the KOI
star and planet.
} \label{Fig:validations}
\end{figure}

We then include both AO data and DSSI speckle data -- we convert the
$K$-band AO data to ${\Delta}Kp$ using the equations of
\citet{howelletal12} while we utilize the 692~nm DSSI data and assume
no difference between this bandpass and $Kp$. This provides a
brightness-dependent limit on the maximum separation between a target
star and a false positive inducing star.  The relative brightnesses
and angular separations of potential background stars excluded by the
photometry, centroid statistics and transit depths for 4 KOIs are
shown in Figure~\ref{Fig:validations}.

There are 18 candidate planets orbiting 12 stars that qualified for
the validation tests.  We use the TRILEGAL galactic simulation
\citep{girardietal12} to first estimate the stellar population within
$1\arcdeg$ around the target star. We then integrate the region of
parameter space not excluded by observations with respect to the
population model. This provides a number of false positive stars,
which is usually much less than 1. We then estimate what fraction of
these are likely to be either background eclipsing binaries or
background planet hosts \citep{slawsonetal11,burkeetal14}.  Finally,
we compare the number of false positives like this in the entire
\kepler data set (i.e. we multiply the number of false positives for
this star by the \kepler sample size of 150,000 \citep{kochetal10}
with the predicted number of planets like this in the \kepler data set
\citep{fressinetal13}).  The ratio of the total planets to the total
number of planets plus false positives yields the probability that a
candidate is a planet.

If the planet candidate is in a multi-planet system we boost the odds
that the candidate is a planet by a factor of $\sim30$ for
two-candidate systems.  This multiplicity boost is justified on the
basis of statistics done on the \kepler sample
\citep{lissaueretal12,lissaueretal14}.  Assuming false positives are
randomly distributed among targets, multiple planet systems should not
have a higher false positive rate than other targets.  It is found
that a large fraction of KOIs with at least one planet candidate are
found to have at least two candidates, meaning a larger fraction of
the planets in these systems are real.  Table~\ref{table:validations}
lists the total validation level for each candidate around an isolated
star for which PRF centroid data is available.  Two of the systems are
designated Kepler-430 and Kepler-431 as newly-validated multi-planet
host stars hosting a total of 5 planets validated at $\geq99.8$\%.
The letter designations for these planets are given in the table.

Several of the planets around these KOIs have previous validation
calculations done or have been deemed validated.  \citet{xie14} showed
that the two current planet candidates of KOI~274 exhibited
anti-correlated transit timing variations, validating both as
interacting planets.  \citet{wangetal14} used the native seeing UKIRT
$J$-band survey images of the \kepler field to help calculate
validation percentages for the multi-planet KOIs~115, 274, 284, 369,
2311 and 3097, including validation boosting due to their planetary
multiplicities.  For KOI~115.01 they report a 99.8\% validation
(considering it part of a 3-planet system in contrast to our study
that excludes KOI~115.03 due to low detection significance).
Wang~et~al. also reported validation levels above 90\% for KOIs~115.02
and 284.01, along with lower levels for the other candidates.
\citet{roweetal14} validated a set of multi-planet KOIs including
KOIs~115, 274, 284, and 369 by incorporating various tests on the
\kepler data and external data products to arrive at $>99$\%
validation levels for their hosted planets.

\section{NEIGHBORING STARS}\label{sec:companions}

\subsection{Observed Properties of Neighboring Stars}
\label{subsec:neighborproperties}

The neighboring point sources (hereafter assumed to be stars) detected
around 7 of the KOIs in the high resolution images as well as native
seeing survey images are listed in Table~\ref{table:secondaries}.  The
table provides the relative separations ($\rho$), position angles
($\theta$), and magnitudes fainter than the KOI (${\Delta}$mag).
Here, each neighbor star is given a designation of ``B'' or ``C'' as
an identifier.  These stars could be foreground or background stars
that are closely aligned by mere chance with the KOI star or may be
gravitationally bound secondaries.  The closer and brighter these
neighboring stars are, the more likely they are to be gravitationally
bound to the KOI, as discussed below.

Table~\ref{table:secondaries} shows that 7 KOI stars have a neighbor
$\leq4\arcsec$ away detected in high resolution images.  In the case
of 5 of these KOIs, one neighbor lies within the exclusion radius for
all planet candidates (KOI~268B, 284B, 1964B, 3255B, and 3284B).  Such
neighbors are potential sources of a false positive (ie. an eclipsing
or transiting system that is blended with the KOI).  Even very faint
neighbors, with magnitudes relative to the KOI star of
${\Delta}Kp=9-10$ can produce a transit-like signal at the level
expected for an Earth-sized planet transiting a Sun-like star
\citep{mortonjohnson11}.

The more distant neighbors should not be considered possible sources
of a false positive.  These stars are nonetheless important to
consider as possible members in a stellar binary with the KOI and for
their dilution of the KOI light curves.  To correct for dilution, a
search was made for any star that could possibly dilute the light
curve at a 1\% level or greater.  Both the high resolution images and
other ground-based imaging surveys such as the UKIRT $J$-band survey
and a catalog of $UBV$ photometry by \citet{everettetal12} were used.
Some neighbors are left out of this search, namely any listed in the
KIC, because any excess flux blended into the light curve by KIC stars
will have already been estimated and removed by the \kepler pipeline.
Two significant KIC stars were noted near the KOIs in this sample:
KIC~11560901, about $8.5\arcsec$ away from KOI~2365 (Kepler-430), and
KIC~8212005, about $12.9\arcsec$ away from KOI~2593.

\subsection{Previous High Resolution Imaging}

Several of these stars have been previously observed with
high-resolution imaging by \citet{adamsetal12,adamsetal13},
\citet{lawetal14} and \citet{lilloboxetal14}.  KOI~1537 was reported
by \citet{adamsetal13} to have a close neighbor with a separation of
$0.13\arcsec$ in $Ks$ AO images.  The same KOI was observed by
\citet{lawetal14} in an optical AO (RoboAO) survey of KOIs, but no
neighbor was found.  Their non-detection would be consistent with the
close separation and RoboAO imaging resolution.  No neighbor of
KOI~1537 is detected in our speckle data.  This is surprising given
the high spatial resolution, which would easily resolve the
separation, and the small magnitude difference ${\Delta}Ks=0.15$
reported by \citet{adamsetal13}.  The limiting contrast for detecting
neighbors at a separation of $0.13\arcsec$ in the 692~nm and 880~nm
images is 5.37 and 4.69 magnitudes respectively.  These apparently
discrepant observations cannot be attributed to a hypothetical very
red neighbor.  The color $880-Ks=1.08$ is found for KOI~1537 from its
isochrone fit, meaning any line-of-sight neighbor would need to be at
least as red as $880-Ks\simeq5.6$, or redder than the reddest stars
($880-Ks=3.9$) found in the isochrones of
\S\ref{subsec:isochronefits}.  In light of this observation, KOI~1537
is treated as a single star here.  \citet{adamsetal12} observed
KOIs~268 and 284 and detected the neighbors of both stars with $J$ and
$K$ magnitude differences in agreement with the values published here.
(Note that the position angles they reported for KOI~268 are
apparently erroneously flipped.)  KOI~115 was observed by both Law et
al. and Lillo-Box et al.  It was seen as single by Law et al., in
agreement with our data.  Lillo-Box et al., who observed with a larger
field-of-view, reported a neighbor to KOI~115 about $4\arcsec$ away
and 8~magnitudes fainter in $i$.  Law et al. and Lillo-Box et al. also
observed KOI~2593.  Both reported it as a single source.  In addition
to KOIs~115, 1537 and 2593, three other stars (KOIs~268, 1964, and
2365) were observed by Law et al. in the optical and, in the case of
KOI~1964, in a near-IR $Ks$ image.  They reported the closest neighbor
to KOI~268 to have an optical magnitude difference of $3.82\pm0.27$,
which is consistent with our near-IR magnitude differences for a
neighbor redder than the KOI.  Their $Ks$ observations of KOI~1964
were in good agreement with ours.  They found KOI~2365 (Kepler-430) to
be single, again in agreement with our results.

\subsection{Distinguishing Bound Companions from Field Stars}\label{subsec:bg_vs_field}

Various evidence may be used to determine if neighboring stars are
gravitationally-bound secondaries or unrelated, line-of-sight field
stars.  A full simulation of the properties and frequency of secondary
stars and field stars could be used.  Instead, in this study, a series
of simpler tests are applied.  These tests consider the brightness of
the neighbor, its angular separation from the KOI star, and the
stellar colors for those stars observed at multiple wavelengths.
Other approaches to using multi-color photometry to investigate the
possible physical association of neighbor stars with KOI stars may be
found in \citet{gillilandetal14} and \citet{lilloboxetal14}.

\subsubsection{Angular Separation and Apparent Brightness}\label{subsubsec:bgprobabilities}

First, the angular separation from the KOI star and magnitude of the
neighbors are examined relative to a random distribution of field
stars at the location of the KOI.  In doing so, an initial assumption
is made that KOI stars are not preferentially co-aligned with
unrelated field stars.  A randomly-generated set of stars representing
a 1~square degree field is produced at the location of the KOI using
the TRILEGAL Galaxy model \citep{girardietal05} Version~1.6 web form.
The model predicts apparent magnitudes in various passbands including
$J$ and $Ks$.  The number of stars in the TRILEGAL model brighter than
the neighbor star is found and multiplied by the ratio of the circular
area inside the neighbor's separation ($\rho$) to the 1~square degree
model field to get a ``background'' probability, $P_{BG}$.  $P_{BG}$
is the likelihood that a field star of the same brightness or brighter
than the neighbor would lie by chance at the same or a smaller angular
separation from the KOI.  In most cases the $Ks$ bandpass is used for
this calculation because it yields the lowest probabilities.  To find
approximate $Ks$ magnitudes for the neighbors, the differential
photometry of the AO images is used along with the $Ks$ magnitude of
the associated 2MASS point source \citep{skrutskieetal06}.  For
neighbor ``C'' of KOI~3284, the $J$ magnitude of the UKIRT survey was
used instead due to unavailable $K$-band data.  The background
probabilities along with the apparent $Ks$ magnitudes derived for each
source are listed in Table~\ref{table:isofits}.  As an empirical check
on this method, the calculation was also run using \kepler magnitudes
of sources extracted from 1~square degree of the KIC at the location
of each KOI.  The probabilities determined from the KIC agreed with
those of the TRILEGAL model to within a factor of 2 (some were higher,
others lower).

Note that in many cases $P_{BG}$ is quite low, a promising indication
that the neighboring stars are gravitationally-bound companions.  The
expectation is that gravitationally-bound secondaries outnumber
co-aligned neighbors in high resolution images such as these,
especially within separations of $\sim1.2\arcsec$ \citep{horchetal14}.
For this reason, these close neighbors are likely dominated by bound
companions.  However, false positive scenarios for small planet
candidate KOIs include the case of large planets transiting background
(or field) stars that are closely co-aligned with the KOIs.  These
cases can appear much like the observed double KOI sources we have
detected in terms of relative magnitude and separation
\citep{fressinetal13}.  For these cases, the assumption that
background stars are randomly-distributed in the sky is invalid and
the low values of $P_{BG}$ are best treated as just one of several
indicators that help distinguish between field stars and bound
companions.  These probabilities are most applicable for neighbor
stars that lie outside the exclusion radius.  On the other hand, a low
value of $P_{BG}$ for a neighbor star inside the exclusion radius can
be explained as either a bound companion or a false positive.

\subsubsection{Colors and Relative Brightness of Neighbors}\label{subsubsec:colormagcomparison}

\begin{figure}
\epsscale{0.8}
\plotone{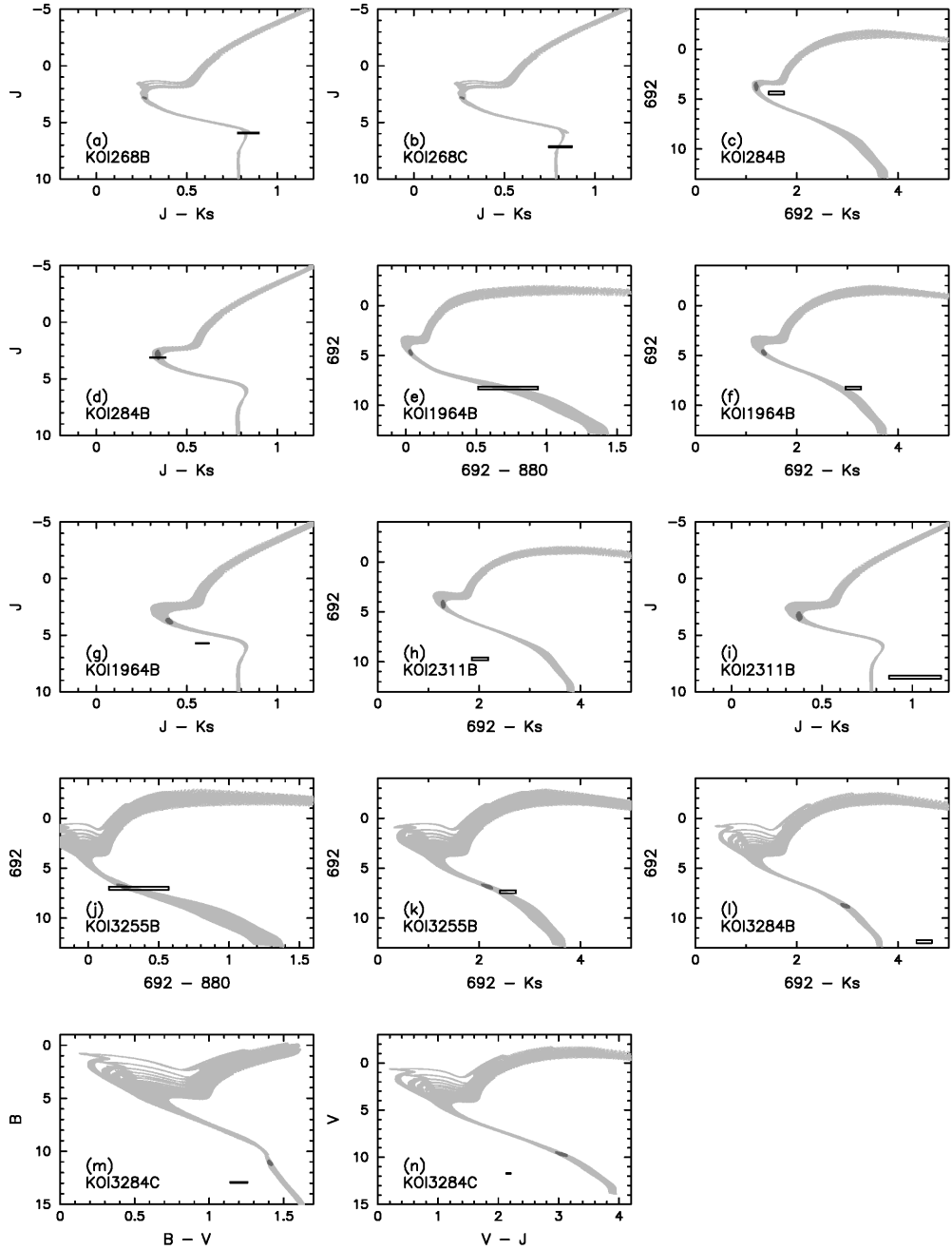}
\caption{
A comparison of the relative magnitudes and colors for 6 KOI stars
with close neighbors.  The absolute magnitudes and colors of each KOI
star are predicted on the basis of fitting its stellar properties
(Table~\ref{table:stellarproperties}) to Dartmouth isochrones.  The
$1\sigma$ confidence intervals are indicated by the dark grey regions
in each plot.  The set of isochrones that pass through this set of
properties are shown in light grey and represent the predicted
absolute magnitudes and colors for any secondary stars.  The
photometry of each neighbor star relative to the KOI star is used to
place it on the same plot.  The location of the neighbor stars are
shown as rectangular boxes that outline the bounds of the $1\sigma$
uncertainties in the relative photometry.  The location of the
neighbor relative to the isochrones is one indicator that helps
distinguish unassociated field stars from gravitationally-bound
companions.
} \label{Fig:isochrones}
\end{figure}

Both stars of gravitationally-bound pairs should lie on the same
isochrone and this can be tested for KOIs that have been observed at
multiple wavelengths.  The test relies on the relative brightnesses
and colors of the two stars.  To determine these, the isochrone fits
are used to find the colors of the KOI stars while the differential
photometry of the imaging provides relative colors and brightnesses of
the neighbor.  Table~\ref{table:neighborcolors} lists the relative
brightnesses and colors for the double KOI sources that have been
observed in more than one filter.  The magnitudes in the table are
absolute magnitudes for the KOI stars and likewise for their neighbor
stars if they are gravitationally bound.  Figure~\ref{Fig:isochrones}
compares the magnitudes and colors of 6 KOI stars and their close
neighbors alongside the isochrones describing the KOI star properties.
The colors and magnitudes of each KOI star (within its $1\sigma$
uncertainty range) are indicated by dark grey regions in each panel.
The light grey regions show the set of isochrones that pass through
these ranges of uncertainty.  The magnitudes in each plot represent
absolute magnitude as predicted by the Dartmouth isochrones or, in the
case of the 692~nm and 880~nm filters, calculated from the isochrones'
SDSS magnitudes as described in \S~\ref{subsec:dssimags}.  The
relative magnitude and colors of the neighbors with respect to the KOI
stars are calculated using the relative photometry provided by the
speckle imaging analysis and near-IR AO images.  The neighboring
stars' colors and magnitudes are shown as rectangular boxes that
indicate the $1\sigma$ photometric uncertainties.

In most panels of Figure~\ref{Fig:isochrones} the relative colors and
brightnesses of the close neighbor stars are consistent with the
isochrones describing the KOI stars.  In other words, most of the
neighbors are consistent with being gravitationally-bound secondaries.
However, in five panels ($g$, $h$, $l$, $m$ and $n$) the neighbor
falls quite far from the isochrones as would be expected in the case
of most field stars.  The neighbors of KOI~1964 (panel $g$) and
KOI~2311 (panel $h$) are fainter and/or bluer in these colors than
main sequence stars at the distance of the KOI, but this situation is
not seen in other plots for the same KOIs (panels $e$, $f$ and $i$).
Neighbor C to KOI~3284 is too faint or blue relative to the Main
Sequence in both colors examined.  In the case of neighbor B to
KOI~3284 (panel $l$), the neighbor is too red relative to the Main
Sequence.  It is also too faint to be a Milky Way giant.  In this
case, the neighbor is so red and faint relative to KOI~3284 (itself a
M dwarf) that, if a bound companion, its luminosity would place it at
an extremely low mass where the model colors are most uncertain.  This
case should be treated with some caution.

\subsubsection{Color Relative to Background Population}

\begin{figure}
\epsscale{0.8}
\plotone{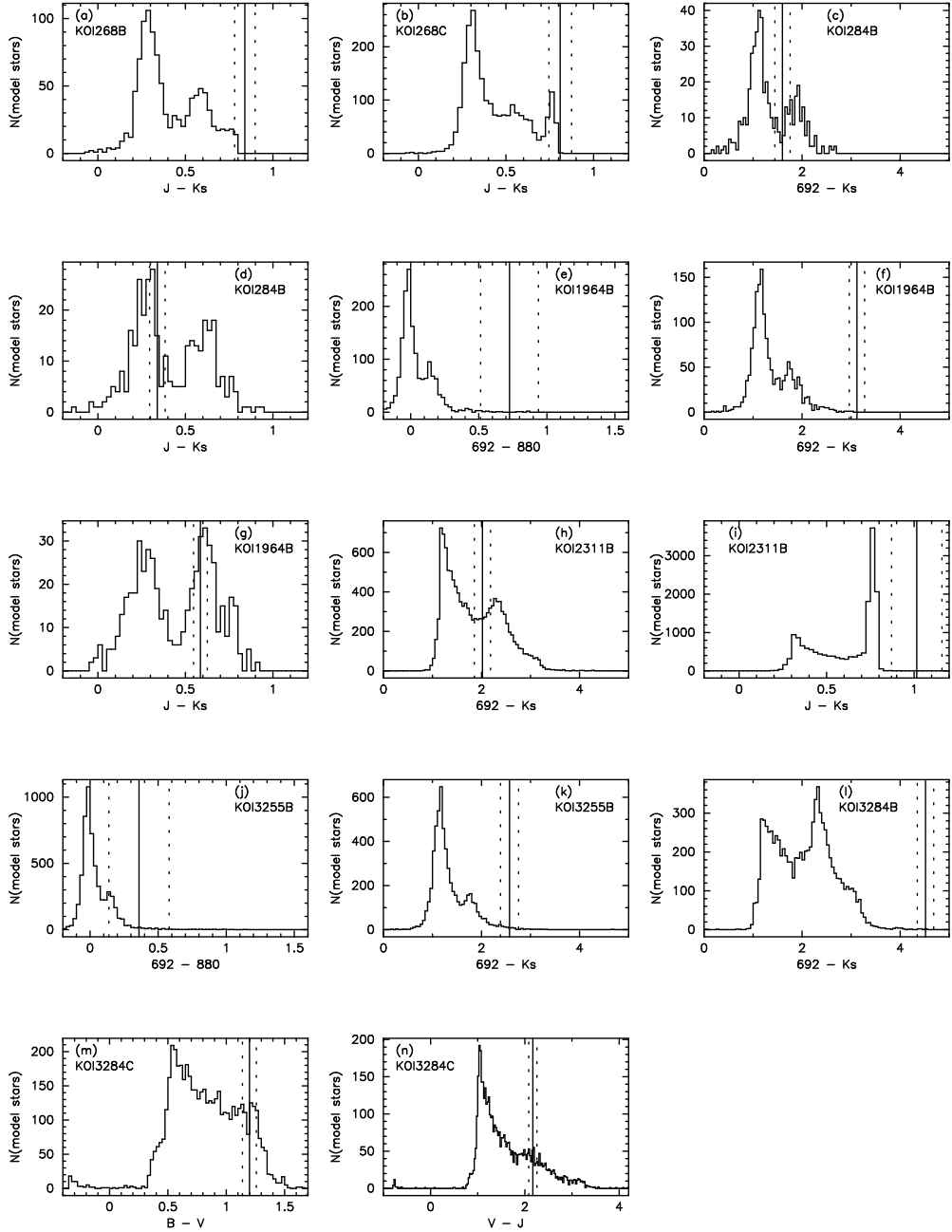}
\caption{
The colors of close neighbors to 6 KOI stars are plotted as solid
vertical lines with their $1\sigma$ uncertainties shown using dotted
vertical lines.  The colors of each neighbor star are determined based
on fitting stellar properties of the KOI star to Dartmouth isochrones
(to predict the KOI magnitudes in each filter) and then applying
offsets in each magnitude based on high-resolution imaging.  The color
distributions for field stars at a similar apparent magnitude as the
neighbor star are calculated using the TRILEGAL Galaxy model at the
location of each KOI and plotted as a histogram in each panel.
} \label{Fig:backgroundcolors}
\end{figure}

To compare the colors of the close neighbors of 6 KOI stars to the
colors of field stars of the same apparent brightness, the TRILEGAL
Galaxy models are used again (the same 1~square degree field
populations used in \S\ref{subsubsec:bgprobabilities}).  This time the
field star populations are restricted to those stars within one
magnitude of the apparent magnitude of the neighbor star in the bluer
of two filters being considered.  The number of field stars is plotted
as a function of colors in Figure~\ref{Fig:backgroundcolors}.  The
colors of the neighboring stars are indicated by vertical lines (solid
lines represent the central value and dotted lines the $1\sigma$
uncertainty interval).

The color distributions of the field stars show several features.  In
$J - Ks$, the peak near 0.25 is due to large numbers of upper main
sequence plus turn-off stars, a second peak near 0.6 is due to
giants, and some plots show a peak at 0.75 due to lower main sequence
dwarfs.  The same features are seen in $692 - Ks$ at 1.1, 1.8, and
2.25 respectively.  The upper Main Sequence plus turn-off stars and
giants show up as peaks near 0 and 0.15 respectively in $692 - 880$.
The red tails of the $692 - 880$ and $692 - Ks$ colors are comprised of
the lowest mass dwarfs.

The field star color distributions are affected by reddening while the
colors derived for the secondary stars are intrinsic colors (zero
reddening effects).  However, extinction is quite small
in the TRILEGAL models where $A_V=0.03-0.04$ to distant lines of sight
in the \kepler field and so reddening corrections in the $692 - Ks$
color would be only $0.02-0.03$~magnitudes using the extinction curve
of \citet{cardellietal89}.  For this reason, no adjustment for the
effects of reddening has been made.

In each panel of Figure~\ref{Fig:backgroundcolors}, the colors of the
neighbor stars are either consistent with the bulk of the field star
distribution or redder than it.  Assuming field stars are distributed
randomly on the sky around each KOI, a first order expectation is that
their colors will be drawn from this same distribution.  For this
reason, the relatively red colors of the neighbors seen in panels $e$,
$f$, $i$, $k$ and $l$ (ie. KOI~1964B, 2311B, 3255B and 3284B) are best
explained by low-mass gravitationally-bound secondaries (although the
extreme color measured for KOI~3284B was difficult to explain as
discussed earlier).  Two other neighbor stars, KOI~268 B and C (panels
$a$ and $b$) are also quite red, but so too are more field stars in $J
- Ks$.

\subsubsection{Assessing the Nature of Each Neighbor Star}

None of the observations definitively distinguishes between
gravitationally-bound or field star neighbors, but in most cases the
evidence points toward the close neighbor being a
gravitationally-bound secondary.  Values of $P_{BG}$ are quite small
for all but the three most distant neighbors (KOI~3255C, KOI~3284C and
KOI~4407C), which means these have a reasonable likelihood of being
nearby field stars.  KOI~1964B and KOI~2311B also show some evidence
of being field stars based on some colors ($J - Ks$ in the case of
KOI~1964B and $692 - Ks$ in the case of KOI~2311B).  However,
KOI~1964B is less conclusive because in the other colors examined it
appears to be a relatively low-mass, red, bound companion.  The photometry
for KOI~3284C is inconsistent with that of a binary companion, but is
internally consistent with that of a background dwarf (see
\S~\ref{subsubsec:fieldstarprops}).

The other 5 neighbors with multi-band photometry (KOIs~268B, 268C,
284B, 3255B and 3284B) are the most consistent with being bound
companions.  KOI3255C has photometry in $J$ only and KOI~4407B and C 
in $Ks$ only, so their natures remain indeterminate.

Overall, among the 11 neighbor stars in this study 8 have multi-band
photometry.  Of those 8, five are deemed likely bound companions, one
a likely field star, and the two others remain too ambiguous to
classify.  Based on this, the fraction of likely bound companions may
be as high as 87.5\% or as low as 62.5\%.  This can be compared to the
lucky imaging survey of 174 candidate or confirmed KOI host stars by
\citet{lilloboxetal14}.  Among their targets observed in both $i$ and
$z$ filters, five were found with close companions, but they
considered only one of them to be a bound companion.  The significance
of the lower fraction of bound companions is difficult to quantify,
but given the larger mean separations for the companions in the
Lillo-Box et al. survey, a greater fraction of field stars is
reasonable.  In a study of 23 KOIs observed in two filters with {\it
  HST/WFC3}, \citet{gillilandetal14} quantified the odds for
neighboring stars to be the bound companions of KOIs.  They found 8
neighboring stars were physically associated with the target KOI, and
6 of these had relatively close separations of $<1\arcsec$.  Clearly,
high resolution imaging inside of $\sim1\arcsec$ is needed to find
most of the wide binary companions to KOIs.

\subsection{Blending Corrections for Crowded KOIs}\label{subsec:neighborsbound}

In order to correct the light curves for the effects of blending,
neighbors' stellar properties must be estimated.  Two separate
scenarios are considered for the status of these neighbor stars: bound
companions and unrelated field stars.  For completeness, and because
most of the neighbor stars are consistent with being
gravitationally-bound secondaries, a calculation is made for each
neighbor assuming it is gravitationally bound.  In the case of
binaries, the secondary star properties are more easily constrained by
the observations because each component shares a common distance,
extinction, composition and age.  The second case, that of neighbor
stars being unrelated field stars, is considered for a few cases where
evidence suggests or allows this scenario.  Constraining the
properties of field stars can prove more difficult.  To find the
stellar properties of assumed secondary stars, the relative photometry
is used for isochrone fits as described in
\S~\ref{subsubsec:secondaryprops}.  To find properties of field star
neighbors, the most likely types of stars are identified in Galaxy
models as discussed in \S~\ref{subsubsec:fieldstarprops}.  For either
case, an aperture correction is found for each star in
\S~\ref{subsubsec:apcorrections}.  Finally, a blending correction, is
formulated in \S~\ref{subsubsec:blendcorrections} based on the $Kp$
magnitudes and aperture corrections for each star in the blend.  These
corrections are used to reevaluate the planet properties for crowded
KOIs as described in \S~\ref{sec:planetproperties}.

\subsubsection{Properties Assuming Neighbors are Secondary Stars}\label{subsubsec:secondaryprops}

\begin{figure}
\epsscale{0.67}
\plotone{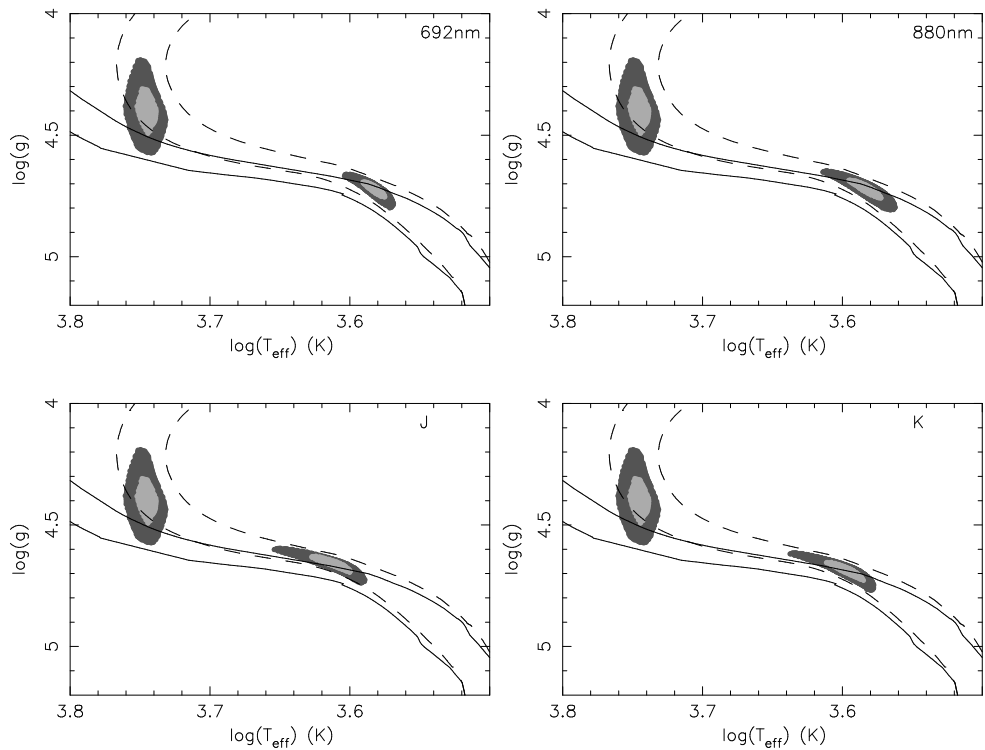}
\caption{
Results of isochrone fits for the double source KOI~1964 (an assumed
binary star).  \teff, \logg\ and \feh\ for the brighter KOI star
(primary) are fitted to a set of Dartmouth isochrones.  The probable
ranges on \logg\ and log(\teff) for the primary are shown in the
shaded region in the upper left of each panel, centered near ${\rm
  log(T_{eff})=3.74}$ and ${\rm log(g)=4.39}$.  The light grey color
indicates the $1\sigma$ range in stellar properties while the darker
grey corresponds to the range $1-2\sigma$.  The same fit also produces
a best value and range for $M_\star$, $L_\star$, $R_\star$ and
absolute magnitudes in various filters.  A neighboring star is found
(secondary) and its properties are found separately in 4 filters (each
shown in one of the 4 panels as labeled by filter).  The data in each
filter consists of a magnitude difference that maps the primary star
properties to a range in secondary properties along the isochrones.
The properties of the secondary are shown near ${\rm
  log(T_{eff})=3.59}$ and ${\rm log(g)=4.70}$ using the same greyscale
representation.  The full range of isochrones are shown as lines in
each panel with dashed lines for 13~Gyr isochrones and solid lines for
1~Gyr isochrones.  The upper lines, dashed or solid, represent ${\rm
  [Fe/H]=+0.4}$ and the lower lines represent ${\rm [Fe/H]=-0.40}$.
See \S\ref{subsec:isochronefits} for details.
} \label{Fig:KOI1964isofit}
\end{figure}

Each image taken in a different filter provides an independent
measurement of the relative brightnesses of the secondary and primary
stars.  The magnitude differences, along with their observational
uncertainties, map the probability distribution of the primary star
properties along a set of isochrones into a distribution of secondary
star properties.  Figure~\ref{Fig:KOI1964isofit} shows the
distribution of primary and secondary star properties in \logg\ and
\teff\ for the double source KOI~1964 observed in four different
filters.  Curves outline the range of isochrones used in the model
fits.  

When relative magnitudes are available in multiple filters, the
secondary properties derived from each are combined in a weighted
average to obtain a final estimate.  One of the most important
secondary properties is $Kp$ because it helps to determine the excess
flux contributed to the light curve by the secondary.
Table~\ref{table:isofits} lists both apparent and absolute $Kp$
magnitude from the isochrone fits for each multiple KOI (apparent $Kp$
are mean values calculated from multiple filters and the absolute $Kp$
values are individual values for each filter).  \kepler magnitudes are
reported in the KIC for each of the blended KOIs considered here.  In
cases where neighbors lie $\leq1.5\arcsec$ from the KOI, the KIC
magnitude is assumed to be the blend of each component whereas more
distant neighbors are assumed to not contribute to the cataloged
magnitude of the KOI.  Table~\ref{table:isofits} also contains the
number of standard deviations each individual filter's results are
from the mean $Kp$.  These numbers indicate how well data from
different passbands match expectations for a bound companion.  The
photometry of both neighbors to KOI~268 and the neighbors of KOIs~284
and 3255 agree well with expectations for bound secondaries.  The
photometry for the neighbor of KOI~1964 agrees less well, but is
plausibly consistent with a bound companion.  The photometry for the
neighbors of KOI~3284 and KOI~2311 is inconsistent with a bound
companion.  This result is similar to the previous analysis indicating
these neighbors are likely to be field stars.
Table~\ref{table:blendedstars} gives the averaged values of
${\Delta}Kp$ for each KOI and neighbor based on the individual
photometric measurements of each system.  This table lists separate
values for different planets in multi-candidate systems since the
blending situation will later be considered separately for each
planet.

\subsubsection{Properties Assuming Neighbors are Field Stars}\label{subsubsec:fieldstarprops}

As discussed in \S~\ref{subsec:bg_vs_field}, there are three
neighboring stars with multi-color photometry that show colors
plausibly inconsistent with those of a gravitationally-bound companion
(KOI~1964B by its $J-Ks$ color, 2311B by its $692-Ks$ color and 3284C
in all colors).  To determine what types of field stars match the
observed brightness and colors, the TRILEGAL Galaxy models are used
once again.  This time the area covered by the Galaxy models is
increased to 4~square degrees for KOI~1964 and 10~square degrees for
KOI~2311 to ensure a rich sample of model stars.  The subset of model
stars whose apparent magnitudes lie within 1~magnitude of the neighbor
star and whose colors fall within the observed uncertainty intervals
listed in Table~\ref{table:neighborcolors} are extracted.  Here, the
apparent magnitude adopted is that of the bluer filter and each subset
contains $\sim200$ stars or more.  The field stars are comprised of
either dwarfs, evolved stars or both in various cases.  For KOI~1964B,
evolved stars (with \logg$<4.0$) are dominant with dwarfs comprising
just 13 out of the 284 total model stars (4.6\%).  For KOI~2311B the
situation is much different with dwarfs accounting for 11771 out of
11825 model stars (99.5\%) and for KOI~3284C the model contained no
evolved stars among the 264 stars matching $B-V$ or the 195 stars
matching in $V-J$.

A similar analysis is done on the three neighbor stars with photometry
available in only one filter.  For these stars, KOI~3255C and
KOI~4407B and C, the subset of stars drawn from the TRILEGAL model is
unconstrained by color and representative of all field stars matching
the brightness of the neighbor.  For KOI~3255C, potential background
stars are almost entirely dwarfs (32158 out of 32724 stars or 98.3\%).
Evolved stars are more likely as background cases for KOI~4407B (only
620 out of 1786 or 34.7\% are dwarfs) and for KOI~4407C (12363 out of
14706 or 84.1\% are dwarfs).

While the stellar properties of the potential background stars can
vary greatly (e.g., different luminosity classes or stellar radii),
the most important property to examine is the relative \kepler
magnitude, which determines the amount of blending.  ${\Delta}Kp$ is
largely a function of effective temperatures and the TRILEGAL stars of
all luminosity classes are combined for its determination.  For cases
where a presumed field star fell within the exclusion radius (field
stars that could be false positive sources), a careful consideration
of its other stellar parameters would be needed, but this situation
was not encountered in our sample.

\begin{figure}
\epsscale{0.67}
\plotone{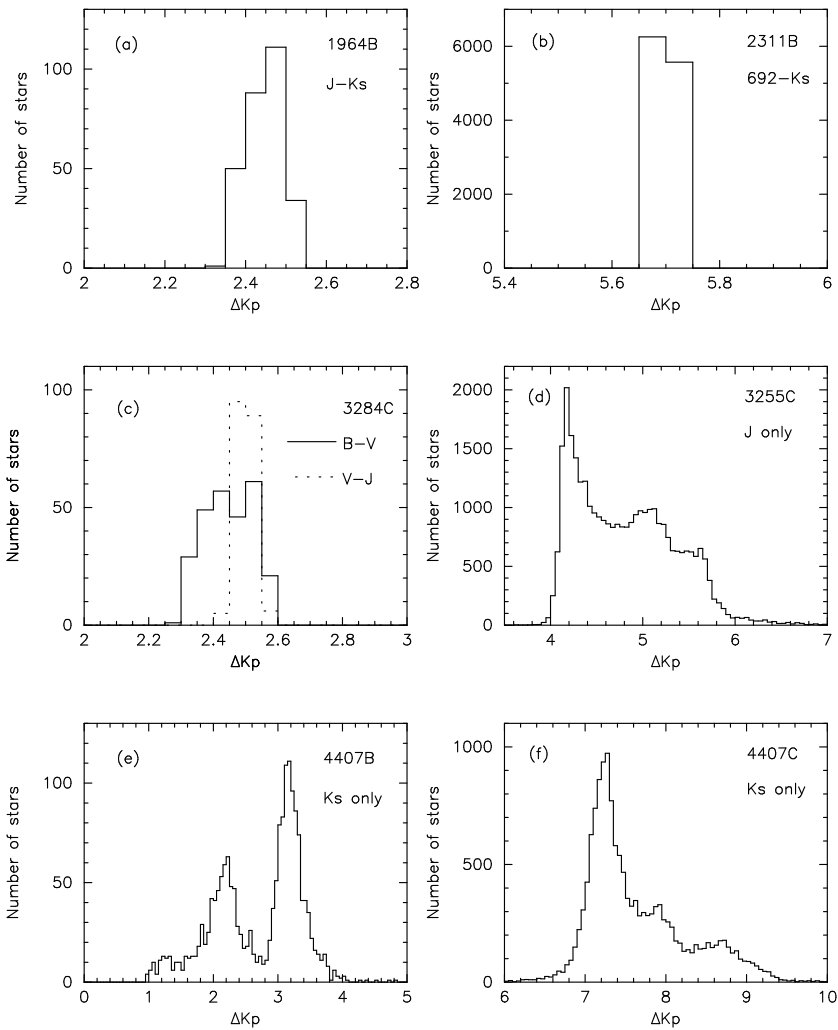}
\caption{
Differences in \kepler magnitudes (${\Delta}Kp$) between 6 KOI stars
and their neighbors based on the TRILEGAL Galaxy models and certain
relative photometry.  This figure assumes the neighbors are field
stars unrelated (not gravitationally-bound) to the the KOI star.  The
histograms show the number of stars matching the neighbor star
brightness and color (when available) as predicted by the TRILEGAL
model as a function of ${\Delta}Kp$.  Each panel is labeled with the
KOI number and the letter designation of the neighbor (B or C).  In
panels $a-c$, multi-band photometry is available to constrain the
color of the neighbor and ${\Delta}Kp$ is confined to a narrow range
of values.  For cases where the neighbor is observed in a single
passband (panels $d-f$), the possible range for ${\Delta}Kp$ is quite
wide ($\sim2$~magnitudes for these near-infrared observations).
} \label{Fig:fieldstarmags}
\end{figure}

Figure~\ref{Fig:fieldstarmags} shows the distributions in ${\Delta}Kp$
for the 6 potential background neighbors and
Table~\ref{table:blendedBGstars} gives the mean differences in \kepler
magnitude for each case.  Note that the table lists the status of two
neighbors as bound companions (KOI~3255B and KOI~3284B) and for these
the \kepler magnitudes are the same as in
Table~\ref{table:blendedstars}.  It is the second neighbors (C) of
these KOIs that are treated as field stars.
Figure~\ref{Fig:fieldstarmags} shows the obvious differences in the
distributions of possible ${\Delta}Kp$ values between those stars with
color information and those without it.  With constraints on the
color, the uncertainty on ${\Delta}Kp$ is as low as
$0.05-0.10$~magnitudes.  This is true even for KOI~1964B where the
color comes from near-IR measurements.  For those neighbors observed
in only one filter, the uncertainty in ${\Delta}Kp$ is at least
2~magnitudes.  This uncertainty is partly due to the near-IR
wavelengths of these observations; a single photometric measurement in
the 692~nm filter, for example, would yield an uncertainty in
${\Delta}Kp$ of $\sim0.2$~magnitudes due to its closer match with the
\kepler bandpass.

\subsubsection{Aperture Corrections}\label{subsubsec:apcorrections}

The fraction of each star's flux that falls in the light curve
aperture is determined using the \kepler Mission data analysis tools
of PyKE \citep{stillbarclay12}, slightly modified for these purposes.
The set of pixels in {\it Kepler}'s pixel light curve files is
analyzed at epochs coinciding with transits calculated from each
candidate's ephemeris.  The average of the effects over all epochs
should represent the effects of blending in the folded light curves
analyzed for planet properties.  For KOIs with orbital periods longer
than 15 days, each transit time is examined.  For shorter orbital
periods, fewer transit times are examined for purposes of efficiency
(e.g., every 2nd or 3rd transit).  \kepler targets shift pixel
location with time, however the most significant differences occur
between different \kepler quarters as the spacecraft rolls by
$90\arcdeg$ and new light curve apertures are used on different CCDs.
Because the KOI stars are the brightest star in each aperture, and
sampled both in the core and the wings, the PyKE tool {\it kepprf} is
used to fit its PRF to the \kepler data and report back a source
center and the fraction of the flux that falls inside the aperture.
For secondary stars, pixel coordinates are fixed relative to the KOI
based on the ground-based astrometry.  The percentage of the pixel
response function of each secondary that falls inside the aperture is
determined and reported in Tables~\ref{table:blendedstars} and
\ref{table:blendedBGstars}.  Where these numbers are reported for
multiple planet candidates of the same KOI, the quite small
differences in flux may be seen.

\subsubsection{Corrections for Blended Light Curves}
\label{subsubsec:blendcorrections}

A blending correction must be made to properly interpret the light
curves of blended KOIs and revise the planet properties derived from
them.  Furthermore, because for some blended KOIs, more than one star
could be the source of the transit-like variations (ie. be the host
star), the effects of blending are considered with respect to each
star.  The goal is to describe an intrinsic light curve for each
possible host star.

The amount of dilution in these light curves and its effect on a key
\kepler measurement, the intrinsic fractional depth of a transit
signal, $\delta_{true}$, may be written as:

\begin{equation}\label{Eq:AA}
\delta_{true,i} = \delta_{obs} \left( 1+\frac{1}{F_i}\sum_{\stackrel{j=1}{j\neq i}}^{N}F_j \right).
\end{equation}

Here, the transit (or eclipse) is assumed to occur to the $i$th star,
and the observed fractional transit depth is $\delta_{obs}$.  The
intrinsic fractional transit depth of the $i$th star,
$\delta_{true,i}$ is found from $\delta_{obs}$ using a blending
correction factor dependent on the flux of all stars in the blend.
Similar treatments were adopted by \citet{lawetal14} and
\citet{lilloboxetal14}.

Tables~\ref{table:blendedstars} and \ref{table:blendedBGstars} list
the blend corrections calculated from Eq.~\ref{Eq:AA}
($\delta_{true}/\delta_{obs}$) using the $Kp$ magnitude for each
component converted to relative flux and with the fluxes modified by
the aperture corrections of \S~\ref{subsubsec:apcorrections}.  It
should be noted that of course these blend corrections are based on
the mean predicted values of ${\Delta}Kp$ and are subject to its
uncertainties which, for cases of field star neighbors observed in a
single filter, can be quite large.

\section{PLANET PROPERTIES}\label{sec:planetproperties}

Revised planet properties are derived based on the new stellar
properties and corrections for blending in the light curves of crowded
KOIs.  All of the KOIs have ``best'' current values for various
stellar and planetary properties, which can be found in the Cumulative
KOI database at the NASA Exoplanet
Archive\footnote{http://exoplanetarchive.ipac.caltech.edu/} or the
stellar properties catalog of \citet{huberetal14}, which includes the
stellar masses.

A first order revised planet radius may be found by scaling the
current value of the planet radius by a factor equal to the ratio of
the new stellar radius to the current stellar radius (one used by the
mission for its current light curve fit).  Such a scaling is ideal for
cases where the revisions to stellar radius are small, although here
it is applied to all cases for illustrative purposes.  For KOIs with
close secondaries, multiple scenarios are considered wherein the
transiting body may orbit the secondary.  In this case, ``close''
secondaries are only those stars falling inside the exclusion radius
around each KOI (so could be the planet host; see \S\ref{sec:fpp}).
Table~\ref{table:planetproperties} lists the isochrone fit values for
stellar radii, masses and effective temperatures for each KOI star as
discussed in \S\ref{sec:stellarproperties}.  In this table, the same
properties listed for the neighbor stars represent those derived from
isochrone fits under the assumption that the neighbors are
gravitationally bound as described in
(\S~\ref{subsubsec:secondaryprops}).  Also listed are transit
(ie. planet orbital) periods, $P_p$, and the planet-to-star radius
ratio, $R_p/R_\star$, a parameter found from light curve fits and
taken from the cumulative KOI database.  This radius ratio, when
multiplied by the new stellar radius, scales the planet radius in
accordance with the revised stellar radius.  The other scaling done to
the planet radii accounts for the blends; here, the value
$\sqrt{\delta_{true}/\delta_{obs}}$, derived from Equation~\ref{Eq:AA}
(see Tables~\ref{table:blendedstars} and \ref{table:blendedBGstars}),
is found and applied as a multiplicative factor that increases planet
radii.  Two (generally) different values of the revised planet radii
are given in Table~\ref{table:planetproperties} representing the case
where each neighbor is assumed to be a binary companion star
($R_{p{\rm BIN}}$) and the case where at least one neighbor is assumed
to be a background star ($R_{p{\rm BG}}$) as listed in
Table~\ref{table:blendedBGstars}.  Cases where the KOI arises due to
transits of an object orbiting a neighboring field star are not
examined because such situations are complicated by large
uncertainties in determining the \kepler magnitude of some neighbors
and the often wide, bimodal distributions in background star radii
(ie. they may be either dwarfs or evolved stars).  Note too that a
complete reevaluation of the planet radii could be performed from new
light curve fits, but such reanalysis lies beyond the scope of this
study.

The magnitude of the ``deblending'' factor varies by KOI.  Consider,
first of all, the KOI star as the host.  For KOI~2311, the secondary
is relatively faint so its effects on the planetary radii are
negligible.  In KOIs~268, 1964, and 4407, estimates for new planet
radii are $1-5$\% higher after deblending.  For KOI~3284, the radius
is 8\% higher and for KOIs~284 and 3255, the planet radii need
adjustment upwards by about 30\%.  In cases where a secondary star is
the potential host star (making the KOI a false positive under our
definition), the effects on planet radius can be even larger and
perhaps be large enough to rule it out as a planet.  This can be seen
in a comparison between the derived radius of the same planet assuming
the KOI as the host star versus the secondary star as the host.  In
the case of KOI~268.01, the radius is 3 times larger if it orbits the
secondary (it is a giant planet rather than sub-Neptune size).  For
KOI~2311.01 and 2311.02, planet sizes change from Earth-like to more
like that of Neptune.  Finally, it is notable that in none of the
cases would a candidate orbiting one of these secondary stars require
a radius exceeding that of a giant planet (ie. require a stellar
eclipse).  These examples make it clear that understanding the role of
light curve blends alongside uncertainties in stellar properties is
vital for understanding the \kepler planet sample.

New stellar masses and effective temperatures invite a recalculation
of planet equilibrium temperatures, an indicator of planet
habitability.  Various assumptions are made in the calculation:
circular orbits and planets with an albedo of 0.3 and uniform surface
temperature.  The planet transit (ie. orbital) periods, $P_P$ are used
to find the semi-major axis, $a$, of each orbit (given the stellar
mass) and this establishes the equilibrium temperatures, $T_{eq}$,
listed in Table~\ref{table:planetproperties}.  Examination of the
equilibrium temperatures reveals the low-$T_{eq}$ candidates selected
for this study.  It also shows that lower values of $T_{eq}$ are
expected for planets orbiting the potential secondaries as opposed to
the KOI stars.

\begin{figure}
\epsscale{0.67}
\plotone{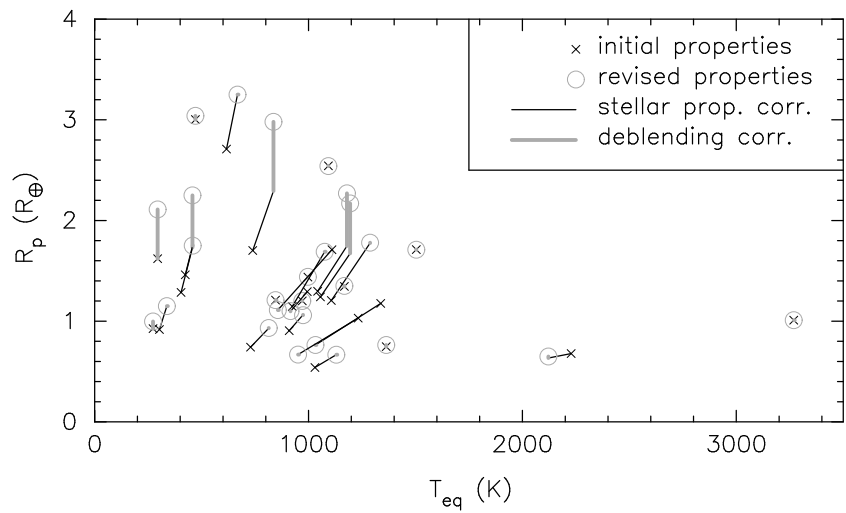}
\caption{
Changes to planet radii ($R_p$) and equilibrium temperatures
($T_{eq}$) made as part of this study.  Here, each planet or candidate
planet is assumed to orbit the KOI star (the brightest star, not any
blended neighbor).  Initial planet properties are shown with lines
connecting them to their revised properties (when applicable) with the
corrections due to revised stellar properties shown separately from
those due to deblending.  Symbols used are defined in the inset box.
} \label{Fig:planetchanges}
\end{figure}

Initial and revised planet radii and equilibrium temperatures are
plotted in Figure~\ref{Fig:planetchanges}.  In this case, each planet
is assumed to orbit the KOI star.  Corrections for the revised stellar
properties and those for deblending are shown separately.  For this
plot, ``initial'' stellar properties are adopted from the catalog of
\citet{huberetal14}, in the case of those stars with KPNO~4m spectra
taken as part of this study, or are the literature values cited in
Table~\ref{table:stellarproperties} otherwise.  The plot shows the
considerable corrections due for some KOIs.  Mostly, the corrections
for stellar properties have resulted in hotter and larger candidates.
Deblending corrections increase planetary size estimates.

\section{INDIVIDUAL KOIs}\label{sec:individualkois}

An overview of new findings for individual planet candidates and
validated or confirmed planets is summarized here.  A complete list of
the host star properties, neighboring stars not listed in the KIC and
planet properties may be found in
Tables~\ref{table:stellarproperties}, \ref{table:secondaries} and
\ref{table:planetproperties} respectively.

\paragraph{KOI~115 (Kepler-105):} Two previously-validated planets 
orbit this isolated solar-type star with orbital periods of 5.412 and
7.126 days.  These planets are validated in our study as well and
given new radii estimates of 2.54 and 1.44~$R_{\bigoplus}$
respectively.
\paragraph{KOI~265:} This isolated, perhaps slightly-evolved, solar-type 
star hosts a single 1.71~$R_{\bigoplus}$ planet candidate with a
3.568-day orbital period.  We present new $J$ and $Ks$ AO imaging in
addition to the speckle imaging for this star to yield a 94\%
validation level.
\paragraph{KOI~268:} This is a 6343~K dwarf star hosting a single
planet candidate in a 110-day orbit.  New $J$ and $K^\prime$ AO
imaging reveals two neighboring stars (B and C) several magnitudes
fainter.  Since neighbor B lies within the exclusion radius, this
planet cannot be validated using our methods.  The proximity and
relative magnitudes in two filters provide evidence that both of these
neighbors are bound companions, but this is less certain than for
other KOIs in our study.  This KOI is subject to slight blending by
its neighbors.  If the candidate orbits the KOI star it is
estimated to have a radius of 3.04~$R_{\bigoplus}$.  If, on the other
hand, it orbits the assumed 4007~K dwarf secondary star (neighbor B), the
radius would be 9.33~$R_{\bigoplus}$ and its equilibrium temperature
would be rather low (217~K).
\paragraph{KOI~274 (Kepler-128):} Two previously-validated planets orbit this
slightly-evolved isolated solar-type star with 15 and 22.8-day orbital
periods.  We validate the planets again with speckle imaging (finding
the host star to be single).  Both planets have radii near
1.2~$R_{\bigoplus}$.
\paragraph{KOI~284 (Kepler-132):} There is a system of three 
previously-validated planets and one planet candidate orbiting this
solar type star, whose stellar properties we characterize with new
spectroscopy.  The \kepler light curve is significantly blended by
flux from a neighboring star, which falls within the exclusion radius
for each planet (so re-validation of this KOI is not attempted here).
In addition to speckle imaging, we publish new $J$ and $Ks$ AO
photometry here.  It is difficult to use the multi-band photometry to
determine if the neighbor is a bound companion or field star.  Its
close angular proximity to the KOI and photometry consistent with that
of a bound companion argue that this is the most likely case.  If a
bound companion star, the planets and candidate planet radii all fall
within a $2-3R_{\bigoplus}$ range and have virtually identical radius
estimates whether they orbit the KOI or secondary star.
\paragraph{KOI~369 (Kepler-144):} This 6157~K dwarf harbors two
previously-validated planets.  We provide new spectroscopic
characterization and $J$-band AO imaging for this star.  No
neighboring stars are found, enabling us to re-validate the planets
based on the speckle and AO imaging.  Kepler-144b and c are found to
have radii of 1.78 and 1.69~$R_{\bigoplus}$ respectively.
\paragraph{KOI~1537:} One candidate planet orbits this 6260~K dwarf with a
period of 10 days.  The speckle imaging presented here shows no
neighboring stars for this KOI in contrast to published AO
observations in $Ks$ \citep{adamsetal13}. The two photometric studies
are apparently irreconcilable assuming a very red star.  Validation is
attempted for this candidate, but the resulting level is fairly low
(86.7\%).  No blending correction is performed for this KOI, resulting
in a planet radius of 1.35~$R_{\bigoplus}$.
\paragraph{KOI~1964:} This 5547~K dwarf is observed with both speckle
imaging and $J$ and $Ks$ AO imaging.  A neighboring star lies about
$0.4\arcsec$ to the north and within the exclusion radius of its
single 2.2-day orbital period planet candidate.  Relative photometry
of the neighboring star results in some ambiguity: near-IR colors
suggest the neighbor is a background field star while the optical
colors are more consistent with those of a bound companion.  Both
cases are examined and the slight blending effects are evaluated and
corrected to conclude that the candidate has a nominal radius of
$0.764$ or $0.785R_{\bigoplus}$ if it orbits the KOI star (the two
cases represent different blending corrections assuming the neighbor
is alternatively a bound companion or field star).  If it orbits the
neighbor and that star is a 3892~K dwarf secondary, the planetary
radius would need to be increased to 2.03~$R_{\bigoplus}$.
\paragraph{KOI~2311:} This solar type dwarf hosts two candidate
planets with orbital periods of approximately 192 and 14
days.  A new spectral characterization is presented for this star.
New optical speckle and AO imaging in $J$ and $K^\prime$ reveal a
faint neighboring star about $1\arcsec$ away.  Since the \kepler data
pipeline did not report the astrometry needed to define an exclusion
radius, the neighbor is also assumed to be the potential host star.
Since the multi-band photometry is ambiguous in terms of determining
if the neighbor star is a bound companion or not, both scenarios are
examined.  The neighbor is relatively faint, so while a deblending
calculation is performed, it has no significant impact on interpreting
the light curve.  The inner planet candidate is found to be
0.932~$R_{\bigoplus}$ if it orbits the KOI and 4.16~$R_{\bigoplus}$ if
it orbits the neighbor (here assuming the bound companion scenario).
The outer planet candidate is cool (337~K) and small
(1.15~$R_{\bigoplus}$) if it orbits the KOI and if it orbits the
neighboring star (again assuming a bound companion), it is cold
(117~K) and larger than Neptune (5.14~$R_{\bigoplus}$).
\paragraph{KOI~2365 (Kepler-430):} This is a solar type dwarf that
hosts two planets and is found to be an isolated star in the speckle
imaging and characterized with a new spectrum.  Both planets are
newly-validated at the 99.9\% level using the speckle data and a
planet multiplicity boost.  The inner planet, Kepler-430b, orbits with
a 36-day period and is found to be 3.25~$R_{\bigoplus}$ with
$T_{eq}=667$~K.  The outer planet, Kepler-430c, orbits in a 111-day
period orbit and is found to be 1.75~$R_{\bigoplus}$ with
$T_{eq}=458$~K.
\paragraph{KOI~2593:} This is an isolated star hosting a single
candidate.  New spectroscopy is presented and shows the KOI to be a
6119~K dwarf.  The planet is validated at a 90.6\% level using speckle
imaging along with a $K^\prime$ AO image and the planet candidate is
found to have a $1.10~R_{\bigoplus}$ radius with an equilibrium
temperature of 915~K.
\paragraph{KOI~2755:} This is an isolated star hosting a single
candidate.  New spectroscopy is presented that shows the KOI to be a
5792~K dwarf.  The planet is validated at a 82.7\% level using speckle
imaging and the planetary properties are found to be
$R_p=1.06R_{\bigoplus}$ and $T_{eq}=974$~K.
\paragraph{KOI~3097 (Kepler-431):} This is a solar type dwarf that
hosts three planets.  A new spectrum is used to characterize the star
as a 6004~K dwarf.  Speckle imaging shows that the star is single and
is used to validate these planets for the first time at the 99.8\%
level.  Each is a small planet orbiting close to the parent star.
Kepler-431b (KOI~3071.02) orbits with a 6.8-day period and is found
with $R_p=0.764$ and $T_{eq}=1032~K$.  Kepler-431c (KOI~3097.03) orbits
with a 8.7-day period and is found with $R_p=0.668$ and
$T_{eq}=951~K$.  Kepler-431d (KOI~3097.01) orbits with a 11.9-day
period and is found to have $R_p=1.11$ and $T_{eq}=856~K$.
\paragraph{KOI~3204:} This is a hot (7338~K) dwarf star with a single
planet candidate in a 0.57-day orbital period.  Speckle imaging shows
this KOI to be a single star, helping confirm the hot planet
candidate's properties, which are calculated here to be
$R_p=1.01R_{\bigoplus}$ and $T_{eq}=3268$~K.  The planet is validated
at a level of 98.5\% based on the speckle images.
\paragraph{KOI~3224:} This is a dwarf slightly cooler than the Sun
(5382~K) as shown by the new spectroscopy in this study.  Speckle
imaging shows it is a single star and validates the planet at a level
of 90.5\%.  Its single planet candidate is sub-Earth size
($0.667~R_{\bigoplus}$) and hot ($T_{eq}=1129~K$).
\paragraph{KOI~3255:} This is a somewhat faint, cool dwarf (4427~K)
that is observed using a combination of optical speckle and $K^\prime$
AO imaging.  It harbors a single, cool planet candidate in a 66.7-day
orbital period.  KOI~3255 has two neighbors: B, a closeby and
relatively bright star at $0.18\arcsec$ separation and a fainter
companion, C, about $3\arcsec$ away that is identified in the UKIRT
$J$-band survey by P. Lucas.  Since neighbor B lies within the
exclusion radius, no validation calculation is done.  Blending effects
are quite significant for this KOI due to the relatively bright
neighbor B.  The multi-band photometry of neighbor B is in agreement
with that of a bound companion as also suggested by its close
separation.  A background star is also a possibility, so both
scenarios are examined.  Neighbor C is observed only in $J$ and has a
relatively wide separation, so its nature is ambiguous.  Given its
relative faintness, however, the lack of color information for
neighbor C is of relatively minor concern.  The candidate planet
radius is found to be $2.11~R_{\bigoplus}$ if it orbits the KOI and
$2.42~R_{\bigoplus}$ if it orbits neighbor B.  Considering the KOI and
(an assumed bound) neighbor B in turn as the potential host star, the
equilibrium temperatures for KOI~3255.01 are quite low, 294~K and
276~K respectively, making it a prime HZ candidate.
\paragraph{KOI~3284:} This is the lowest mass KOI star in the study,
a 3688~K dwarf.  It harbors a single planet candidate with a 35-day
orbital period.  It is observed using $K^\prime$ AO along with optical
speckle imaging and found to have a nearby companion, B, at
$0.44\arcsec$ separation.  Another neighbor, C, is found in $UBV$ and
$J$-band surveys of the \kepler field.  Neighbor B falls within the
exclusion radius so no validation is attempted.  Blending effects are
fairly significant for this KOI.  Once corrected, the planet candidate
is found to have nominally $R_p=0.99$ (assuming background neighbors)
or $1.00R_\bigoplus$ (assuming bound neighbors) and $T_{eq}=272~K$ if
it orbits the KOI star.  If it orbits neighbor B, and this neighbor is
a bound companion, one finds $R_p=1.46R_{\bigoplus}$ and
$T_{eq}=184~K$.  The photometry for neighbor C is best explained as
that of a background dwarf.  KOI~3284.01 is a small HZ candidate.
\paragraph{KOI~4407:} This is a 6408~K dwarf as found in the new
spectral characterization of this study.  It hosts a single small
planet candidate with a 1.34-day orbital period.  The star was
observed using both speckle and $Ks$-band AO imaging and found to have
two neighbors as observed only within the wider field of the AO
images.  These neighbors both lie outside of the exclusion radius and
validation is possible at a 19.2\% level (the 17~ppm light curve
transit depth is exceptionally shallow making validation difficult).
Since the neighboring stars are observed in a single passband, their
nature (whether secondary/tertiary or field stars) is ambiguous.
There is some light curve blending due to the neighbors.  If both
neighbors are gravitationally bound to the KOI, the planetary
properties would be $R_p=0.65R_\bigoplus$ and $T_{eq}=2121~K$.
Nominally, the radius would be very similar if the neighbors are field
stars, but with additional uncertainty because ${\Delta}Kp$ is poorly
constrained.  While nominally, ${\Delta}Kp = 2.70$ and 7.64 for field
star neighbors B and C respectively, their relative ${\Delta}Kp$
magnitudes could be as bright as 1.0 and 6.6 or as faint as 4.0 and
9.4 respectively (see Figure~\ref{Fig:fieldstarmags}).  In such
eventualities, the blending-corrected planet radius could range from
$R_p=0.64$ (for the case of faintest possible neighbors) to 0.75
(assuming each neighbor is as bright as possible).

\section{CONCLUSION}\label{sec:conclusion}

A high resolution speckle imaging survey was done on 18 KOI stars that
host a total of 28 planets and candidate planets.  This was
supplemented by near-IR adaptive optics imagery of 10 and new
spectroscopic characterizations for 11 of these stars.  Validations
(planet status confidence levels) are calculated for 18 of the planets
or candidate planets that orbit the 12 host stars that are
sufficiently isolated from detectable neighbors.  There are 12 of the
18 planets validated at levels $>98$\%.  Five of these planets are
first time validations with levels of at least $99.8$\% (validating
the two-planet system KOI~2365 as Kepler-430 and three-planet system
KOI~3097 as Kepler-431).  The high resolution imaging helped discover
and then provide multi-color photometry to characterize close neighbor
stars to 7 of the KOIs.  These data, along with stellar
characterization of the primary stars, were used to examine the
relationship of the neighbors to the KOIs (gravitationally bound
vs. field stars), and ``deblend'' the light curves by removing the
excess light curve dilution due to neighbor stars.  A reevaluation of
the planet properties was done for the KOIs, accounting for revised
host star properties and blending effects.  Potential cases where
neighboring stars could be the source of false positive planet signals
were also evaluated.

Further observations can help to solve some of the unresolved
questions surrounding KOIs with neighboring stars.  As shown, for the
example targets, when double KOIs are observed in two or more filters,
it is much easier to characterize both stars.  Because much data is
already published or publicly available, including good stellar
characteristics for KOIs and imaging at multiple wavelengths (e.g.,
wide-field surveys of the \kepler field in addition to targeted
surveys of KOIs), it should be possible to apply many of the methods
from this study to larger numbers of stars and determine,
statistically, how blended KOIs bias the \kepler planet sample.  By
quantifying the biases, appropriate corrections can be applied.

There are uses for compiling a list of validated or candidate planets
harbored by binary stars.  Many stars are binary, but how does that
environment affect planet occurrence and orbital properties?  To
better discriminate binary companions from unrelated, closely
co-aligned field stars, repeated speckle astrometry can be used to
find common proper motions pairs.  For many KOIs, this will be a
straightforward test.  For example, the KIC lists proper motion
measurements for 6 of the 18 stars in our sample with values ranging
from $4-24$~mas~yr$^{-1}$.  Two of the KOIs found to be double are
among this group and have proper motions of 4 and 6~mas~yr$^{-1}$.  As
\citet{horchetal12} have shown, and with some more analysis of recent
data, speckle imaging at Gemini yields astrometric precision between
$1-1.5$~mas for targets in the brightness range of KOIs.  At the
observed rates, relative proper motions (or common proper motions) can
be detected using a pair of observations spaced 1 or 2 years apart.
Recent work by \citet{benedictetal14} shows that proper motions can now
be derived for KOIs based on \kepler pixel data, yielding precision 3
times better than in the existing catalogs.  This level of astrometry
could prove very helpful in identifying common proper motions.
Additionally, ESA's {\it Gaia} Mission \citep{perrymanetal01} promises
to deliver a revolutionary astrometric dataset, impacting many fields.
Once available, {\it Gaia} data should result in the reevaluation of
\kepler data, including the definitive detection of many astrometric
binaries.
 
Most of the data presented here is made available to the community for
download at the \kepler Community Follow-up Observing Program website
(CFOP)$^8$, a service of the NASA Exoplanet Archive.  These data
include tabulated sensitivity curves for each of the speckle
observations.

\acknowledgments

The authors acknowledge the support of many people and programs that
made this work possible.  This paper includes data collected by the
\kepler Mission. Funding for the mission is provided by the NASA
Science Mission directorate.  M. E. Everett received support through
NASA Agreement NNX13AB60A.  T. Barclay was partially supported by a
NASA Keck PI Data Award, administered by the NASA Exoplanet Science
Institute.  Comments to improve upon a draft of this paper were
received from an anonymous referee, who we thank for the help.

Data for this paper were obtained from numerous sources, including:
(1) The NASA Exoplanet Archive, which is operated by the California
Institute of Technology, under contract with the National Aeronautics
and Space Administration under the Exoplanet Exploration Program; (2)
the Mikulski Archive for Space Telescopes (MAST). STScI is operated by
the Association of Universities for Research in Astronomy, Inc., under
NASA contract NAS5-26555. Support for MAST for non-HST data is
provided by the NASA Office of Space Science via grant NNX13AC07G and
by other grants and contracts; and (3) the Two Micron All Sky Survey,
which is a joint project of the University of Massachusetts and the
Infrared Processing and Analysis Center/California Institute of
Technology, funded by the National Aeronautics and Space
Administration and the National Science Foundation.  This work also
made use of PyKE \citep{stillbarclay12}, a software package for the
reduction and analysis of Kepler data. This open source software
project is developed and distributed by the NASA Kepler Guest Observer
Office.

The speckle imaging observations were obtained as part of the program
GN-2013B-Q-87 at the Gemini Observatory, which is operated by the
Association of Universities for Research in Astronomy, Inc., under a
cooperative agreement with the NSF on behalf of the Gemini
partnership: the National Science Foundation (United States), the
National Research Council (Canada), CONICYT (Chile), the Australian
Research Council (Australia), Minist\'{e}rio da Ci\^{e}ncia,
Tecnologia e Inova\c{c}\~{a}o (Brazil) and Ministerio de Ciencia,
Tecnolog\'{i}a e Innovaci\'{o}n Productiva (Argentina).  We are very
grateful for the excellent support of the Gemini administration and
support staff who helped make the visiting instrument program possible
and the DSSI observing run a great success.

Some of the data presented herein were obtained at the W.M. Keck
Observatory, which is operated as a scientific partnership among the
California Institute of Technology, the University of California and
the National Aeronautics and Space Administration. The Observatory was
made possible by the generous financial support of the W.M. Keck
Foundation.

Finally, the authors wish to recognize and acknowledge the very
significant cultural role and reverence that the summit of Mauna Kea
has always had within the indigenous Hawaiian community.  We are most
fortunate to have the opportunity to conduct observations from this
mountain.

Facilities: \facility{Gemini:Gillett}, \facility{Mayall}, 
\facility{Keck:II}, \facility{Hale}, \facility{Shane}, \facility{Kepler}

\newpage

\clearpage

\clearpage

\clearpage

\clearpage

\clearpage

\clearpage

\clearpage

\clearpage

\clearpage

\begin{deluxetable}{clrccc}
\tabletypesize{\scriptsize}
\tablecolumns{10}
\tablewidth{0pt}
\tablecaption{Speckle Imaging Observations\label{table:spklobservations}}
\tablehead{
  \colhead{KOI} & \colhead{Kepler Name} & \colhead{KIC ID} & \colhead{Kepler Mag.} & \colhead{Date} &
  \colhead{Number of 60ms frames} \\
  \colhead{} & \colhead{} & \colhead{} & \colhead{} & \colhead{(UT)} &
  \colhead{(thousands)} \\
}
\startdata
115  & Kepler-105 & 9579641  & 12.791 & 2013 July 25 & 12  \\
265  & \nodata    & 12024120 & 11.994 & 2013 July 28 & 9   \\
268  & \nodata    & 3425851  & 10.560 & 2013 July 25 & 3   \\
274  & Kepler-128 & 8077137  & 11.390 & 2013 July 27 & 6   \\
284  & Kepler-132 & 6021275  & 11.818 & 2013 July 25 & 6   \\
369  & Kepler-144 & 7175184  & 11.992 & 2013 July 27 & 9   \\
1537 & \nodata    & 9872292  & 11.740 & 2013 July 27 & 6   \\
1964 & \nodata    & 7887791  & 10.687 & 2013 July 27 & 3   \\
2311 & \nodata    & 4247991  & 12.570 & 2013 July 25 & 9   \\
2365 & Kepler-430 & 11560897 & 13.848 & 2013 July 25 & 18  \\
2593 & \nodata    & 8212002  & 11.714 & 2013 July 27 & 6   \\
2755 & \nodata    & 3545135  & 12.147 & 2013 July 27 & 9   \\
3097 & Kepler-431 & 7582689  & 11.973 & 2013 July 27 & 6   \\
3204 & \nodata    & 11456279 & 11.825 & 2013 July 27 & 6   \\
3224 & \nodata    & 10384298 & 12.192 & 2013 July 27 & 9   \\
3255 & \nodata    & 8183288  & 14.352 & 2013 July 27 & 21  \\
3284 & \nodata    & 6497146  & 14.467 & 2013 July 25 & 25  \\
4407 & \nodata    & 8396660  & 11.179 & 2013 July 28 & 6   \\
\enddata
\end{deluxetable}

\begin{deluxetable}{crclcc}
\tabletypesize{\scriptsize}
\tablecolumns{6}
\tablewidth{0pt}
\tablecaption{Near-infrared Adaptive Optics Observations\label{table:AOobservations}}
\tablehead{
  \colhead{KOI} & \colhead{Kepler Name} & \colhead{Kepler Mag.} & \colhead{Date} &
  \colhead{Telescope/Instrument\tablenotemark{a}} & \colhead{Filter}  \\
  \colhead{} & \colhead{} & \colhead{} & \colhead{(UT)} &
  \colhead{} & \colhead{} \\
}
\startdata
265  & \nodata    & 11.994 & 2010 July 2   & Palomar/Pharo & $J,Ks$       \\
268  & \nodata    & 10.560 & 2012 July 4   & Keck/NIRC2    & $J,K^\prime$  \\
284  & Kepler-132 & 11.818 & 2010 July 1-2 & Palomar/Pharo & $J,Ks$       \\
369  & Kepler-144 & 11.992 & 2011 Sept. 10 & Lick/IRCAL    & $J$          \\
1964 & \nodata    & 10.687 & 2013 June 26  & Palomar/Pharo & $J,Ks$       \\
2311 & \nodata    & 12.570 & 2012 Aug. 25  & Keck/NIRC2    & $J,K^\prime$  \\
2593 & \nodata    & 11.714 & 2013 July 7   & Keck/NIRC2    & $K^\prime$    \\
3255 & \nodata    & 14.352 & 2012 Aug. 25  & Keck/NIRC2    & $K^\prime$    \\
3284 & \nodata    & 14.467 & 2013 July 6   & Keck/NIRC2    & $K^\prime$    \\
4407 & \nodata    & 11.179 & 2013 June 27  & Palomar/Pharo & $Ks$         \\
\enddata
\tablenotetext{a}{Palomar indicates Hale~5m, Keck indicates Keck-II~10m and
			  Lick indicates Shane~3.5m telescopes.}
\end{deluxetable}

\begin{deluxetable}{cccc}
\tabletypesize{\scriptsize}
\tablecolumns{10}
\tablewidth{0pt}
\tablecaption{Spectroscopy Observations\label{table:spectraobs}}
\tablehead{
  \colhead{KOI} & \colhead{Kepler Name} & \colhead{Kepler Mag.} & \colhead{Date Observed} \\
  \colhead{} & \colhead{} & \colhead{} & \colhead{(UT)} \\
}
\startdata
115  & Kepler-105  & 12.791 & 2010 May 24   \\
265  & \nodata     & 11.994 & 2010 Sept. 14 \\
284  & Kepler-132  & 11.818 & 2013 Sept. 1  \\
369  & Kepler-144  & 11.992 & 2013 Sept. 1  \\
2311 & \nodata     & 12.570 & 2013 Sept. 1  \\
2365 & Kepler-430  & 13.848 & 2013 Sept. 1  \\
2593 & \nodata     & 11.714 & 2013 Sept. 1  \\
2755 & \nodata     & 12.147 & 2013 Sept. 1  \\
3097 & Kepler-431  & 11.973 & 2013 Sept. 1  \\
3224 & \nodata     & 12.192 & 2013 Sept. 1  \\
4407 & \nodata     & 11.179 & 2013 Sept. 1  \\
\enddata
\end{deluxetable}

\begin{deluxetable}{clcccccc}
\tabletypesize{\scriptsize}
\tablecolumns{8}
\tablewidth{0pt}
\tablecaption{Stellar Properties\label{table:stellarproperties}}
\tablehead{
  \colhead{KOI} & \colhead{Kepler Name} & \colhead{$T_{\rm eff}$} & \colhead{log(g)} & \colhead{[Fe/H]} &
  \colhead{$R_\star$} & \colhead{$M_\star$} & \colhead{Reference\tablenotemark{a}} \\
  \colhead{} & \colhead{} & \colhead{(K)} & \colhead{(cgs)} & \colhead{(dex)} &
  \colhead{($R_\odot$)} & \colhead{($M_\odot$)} & \colhead{} \\
}
\startdata
115 & Kepler-105 & $6065\pm75$ & $4.43\pm0.15$ & $-0.10\pm0.10$ & $1.015\begin{array}{c}+0.189\\-0.071\end{array}$ & $1.027\begin{array}{c}+0.057\\-0.035\end{array}$ & 1 \\[10pt]
265 & \nodata & $5915\pm75$ & $4.07\pm0.15$ & $0.06\pm0.10$ & $1.564\begin{array}{c}+0.456\\-0.252\end{array}$ & $1.097\begin{array}{c}+0.122\\-0.054\end{array}$ & 1 \\[10pt]
268 & \nodata & $6343\pm85$ & $4.259\pm0.010$ & $-0.040\pm0.101$ & $1.366\pm0.026$ & $1.230\pm0.058$ & 2 \\[10pt]
274 & Kepler-128 & $6072\pm75$ & $4.070\pm0.011$ & $-0.090\pm0.101$ & $1.659\pm0.038$ & $1.184\pm0.074$ & 2 \\[10pt]
284 & Kepler-132 & $5879\pm75$ & $4.15\pm0.15$ & $-0.04\pm0.10$ & $1.408\begin{array}{c}+0.284\\-0.240\end{array}$ & $1.023\begin{array}{c}+0.080\\-0.055\end{array}$ & 3 \\[10pt]
369 & Kepler-144 & $6157\pm75$ & $4.14\pm0.15$ & $-0.02\pm0.10$ & $1.491\begin{array}{c}+0.288\\-0.247\end{array}$ & $1.126\begin{array}{c}+0.108\\-0.049\end{array}$ & 3 \\[10pt]
1537 & \nodata & $6260\pm116$ & $4.047\pm0.014$ & $0.100\pm0.109$ & $1.824\pm0.049$ & $1.366\pm0.101$ & 2 \\[10pt]
1964 & \nodata & $5547\begin{array}{c}+109\\-91\end{array}$ & $4.388\begin{array}{c}+0.107\\-0.126\end{array}$ & $-0.040\begin{array}{c}+0.160\\-0.140\end{array}$ & $0.989\begin{array}{c}+0.177\\-0.109\end{array}$ & $0.871\begin{array}{c}+0.068\\-0.038\end{array}$ & 4 \\[10pt]
2311 & \nodata & $5657\pm75$ & $4.29\pm0.15$ & $0.15\pm0.10$ & $1.182\begin{array}{c}+0.220\\-0.195\end{array}$ & $0.975\begin{array}{c}+0.046\\-0.039\end{array}$ & 3 \\[10pt]
2365 & Kepler-430 & $5884\pm75$ & $4.15\pm0.15$ & $0.20\pm0.10$ & $1.485\begin{array}{c}+0.266\\-0.234\end{array}$ & $1.166\begin{array}{c}+0.134\\-0.095\end{array}$ & 3 \\[10pt]
2593 & \nodata & $6119\pm75$ & $4.21\pm0.15$ & $0.16\pm0.10$ & $1.453\begin{array}{c}+0.304\\-0.287\end{array}$ & $1.230\begin{array}{c}+0.091\\-0.093\end{array}$ & 3 \\[10pt]
2755 & \nodata & $5792\pm75$ & $4.29\pm0.15$ & $0.01\pm0.10$ & $1.172\begin{array}{c}+0.236\\-0.173\end{array}$ & $0.973\begin{array}{c}+0.047\\-0.037\end{array}$ & 3 \\[10pt]
3097 & Kepler-431 & $6004\pm75$ & $4.40\pm0.15$ & $0.07\pm0.10$ & $1.092\begin{array}{c}+0.191\\-0.109\end{array}$ & $1.071\begin{array}{c}+0.059\\-0.037\end{array}$ & 3 \\[10pt]
3204 & \nodata & $7338\begin{array}{c}+226\\-336\end{array}$ & $4.225\begin{array}{c}+0.060\\-0.445\end{array}$ & $0.070\begin{array}{c}+0.170\\-0.390\end{array}$ & $1.593\begin{array}{c}+1.273\\-0.202\end{array}$ & $1.553\begin{array}{c}+0.375\\-0.225\end{array}$ & 5 \\[10pt]
3224 & \nodata & $5382\pm75$ & $4.30\pm0.15$ & $0.10\pm0.10$ & $0.962\begin{array}{c}+0.100\\-0.091\end{array}$ & $0.866\begin{array}{c}+0.040\\-0.021\end{array}$ & 3 \\[10pt]
3255 & \nodata & $4427\begin{array}{c}+133\\-129\end{array}$ & $4.639\begin{array}{c}+0.055\\-0.033\end{array}$ & $-0.320\begin{array}{c}+0.340\\-0.320\end{array}$ & $0.622\begin{array}{c}+0.056\\-0.060\end{array}$ & $0.615\begin{array}{c}+0.066\\-0.049\end{array}$ & 6 \\[10pt]
3284 & \nodata & $3688\begin{array}{c}+73\\-50\end{array}$ & $4.788\begin{array}{c}+0.060\\-0.080\end{array}$ & $-0.100\pm0.100$ & $0.463\begin{array}{c}+0.070\\-0.050\end{array}$ & $0.479\begin{array}{c}+0.060\\-0.050\end{array}$ & 7 \\[10pt]
4407 & \nodata & $6408\pm75$ & $4.22\pm0.15$ & $0.01\pm0.10$ & $1.435\begin{array}{c}+0.329\\-0.265\end{array}$ & $1.234\begin{array}{c}+0.102\\-0.065\end{array}$ & 3 \\[10pt]

\enddata
\tablenotetext{a}{
1 = $T_{\rm eff}$, log(g) and [Fe/H] from \citet{everettetal13} and $R_\star$ and $M_\star$ from this work;
2 = All values from the stellar properties catalog (SPC) of \citet{huberetal14} based on data from \citet{huberetal13};
3 = Stellar properties all from this work;
4 = $T_{\rm eff}$, log(g) and [Fe/H] are SPC values from \citet{huberetal14} based on data from \citet{batalhaetal13}
    and $R_\star$ and $M_\star$ are from this work;
5 = All values from the SPC of \citet{huberetal14};
6 = $T_{\rm eff}$, log(g) and [Fe/H] are SPC values from \citet{huberetal14} based on data from \citet{pinsonneaultetal12}
    and $R_\star$ and $M_\star$ are from this work;
7 = $T_{\rm eff}$, log(g) and [Fe/H] are SPC values from \citet{huberetal14} based on data from \citet{dressingcharbonneau13}
    and $R_\star$ and $M_\star$ are from this work}
\end{deluxetable}

\begin{deluxetable}{rlrll}
\tabletypesize{\scriptsize}
\tablecolumns{5}
\tablewidth{0pt}
\tablecaption{Planet Validation Results\label{table:validations}}
\tablehead{
  \colhead{KOI} & \colhead{Kepler Name} & \colhead{${\Delta}Kp_{max}$} & \colhead{$r_{ex}$} &
  \colhead{validation level} \\
  \colhead{} & \colhead{} & \colhead{(mag.)} & \colhead{($\arcsec$)} & \colhead{} \\
}
\startdata
115.01 & Kepler-105b & 7.31 & 0.195 & 0.99996 \\
115.02 & Kepler-105c & 8.58 & 0.72 & 0.9997 \\
265.01 & \nodata & 9.29 & 0.99 & 0.940 \\
268.01 & \nodata & 7.51 & 2.31 & \nodata \\
274.01 & Kepler-128b & 10.00 & 1.56 & 0.998 \\
274.02 & Kepler-128c & 9.96 & 2.1 & 0.998 \\
284.01 & Kepler-132d & 8.70 & 1.08 & \nodata \\
284.02 & Kepler-132c & 9.03 & 1.44 & \nodata \\
284.03 & Kepler-132b & 9.15 & 1.53 & \nodata \\
284.04 & \nodata & 8.95 & 7.8 & \nodata \\
369.01 & Kepler-144b & 9.01 & 1.29 & 0.989 \\
369.02 & Kepler-144c & 9.10 & 0.9 & 0.994 \\
1537.01 & \nodata & 9.72 & 1.38 & 0.867 \\
1964.01 & \nodata & 9.96 & 1.86 & \nodata \\
2311.01 & \nodata & 9.18 & \nodata & \nodata \\
2311.02 & \nodata & 10.27 & \nodata & \nodata \\
2365.01 & Kepler-430b & 8.18 & 0.84 & 0.9993 \\
2365.02 & Kepler-430c & 8.88 & 2.64 & 0.999 \\
2593.01 & \nodata & 9.84 & 3.3 & 0.906 \\
2755.01 & \nodata & 9.56 & 0.96 & 0.827 \\
3097.01 & Kepler-431d & 9.90 & 1.23 & 0.9994 \\
3097.02 & Kepler-431b & 10.48 & 1.32 & 0.998 \\
3097.03 & Kepler-431c & 10.55 & 2.25 & 0.998 \\
3204.01 & \nodata & 10.42 & 0.36 & 0.985 \\
3224.01 & \nodata & 10.03 & 2.67 & 0.905 \\
3255.01 & \nodata & 7.45 & 0.57 & \nodata \\
3284.01 & \nodata & 7.79 & 1.41 & \nodata \\
4407.01 & \nodata & 11.20 & 0.78 & 0.192 \\
\enddata
\end{deluxetable}

\begin{deluxetable}{cccccccc}
\tabletypesize{\scriptsize}
\tablecolumns{10}
\tablewidth{0pt}
\tablecaption{Neighboring Stars Not listed in the Kepler Input Catalog\label{table:secondaries}}
\tablehead{
  \colhead{KOI} & \colhead{KIC} & \colhead{Source\tablenotemark{a}} & \colhead{Filter} & \colhead{Star} &
  \colhead{$\theta$ ($\arcdeg$)} & \colhead{$\rho$ ($\arcsec$)} &  \colhead{$\Delta$mag.} \\
}
\startdata
268  &  3425851 & NIRC2 & $J$       & B & $267.69\pm0.02$\tablenotemark{b} & $1.7591\pm0.0002$\tablenotemark{b} & $3.11\pm0.05$  \\
     &          & NIRC2 & $K^\prime$ & B & $267.69\pm0.02$\tablenotemark{b} & $1.7591\pm0.0002$\tablenotemark{b} & $2.54\pm0.03$  \\ 
     &          & NIRC2 & $J$       & C & $310.19\pm0.02$\tablenotemark{b} & $2.5243\pm0.0006$\tablenotemark{b} & $4.33\pm0.05$  \\ 
     &          & NIRC2 & $K^\prime$ & C & $310.19\pm0.02$\tablenotemark{b} & $2.5243\pm0.0006$\tablenotemark{b} & $3.79\pm0.04$  \\ 

284\tablenotemark{c}  &  6021275 & DSSI  & 692nm     & B &  97.44   & 0.8672  & $0.66\pm0.15$    \\
     &          & DSSI  & 880nm     & B &  97.25   & 0.8681  & \nodata \\
     &          & Pharo & $J$       & B & \nodata  & \nodata & 0.26    \\
     &          & Pharo & $Ks$      & B & \nodata  & \nodata & 0.26    \\

1964 &  7887791 & DSSI  & 692nm     & B &   1.72   & 0.3916  & $3.54\pm0.15$    \\
     &          & DSSI  & 880nm     & B &   2.81   & 0.4039  & $2.85\pm0.15$    \\
     &          & Pharo & $J$       & B & \nodata  & \nodata & 1.96    \\
     &          & Pharo & $Ks$      & B & \nodata  & \nodata & 1.78    \\

2311 &  4247991 & DSSI  & 692nm     & B &  69.03   & 1.0295  & $5.47\pm0.15$    \\
     &          & NIRC2 & $J$       & B &  $70.19\pm0.04$\tablenotemark{b} & $1.0264\pm0.0003$\tablenotemark{b} & $5.38\pm0.13$  \\
     &          & NIRC2 & $K^\prime$ & B &  $70.19\pm0.04$\tablenotemark{b} & $1.0264\pm0.0003$\tablenotemark{b} & $4.74\pm0.06$  \\

3255 &  8183288 & DSSI  & 692nm     & B & 336.41   & 0.1812  & $0.52\pm0.15$    \\
     &          & DSSI  & 880nm     & B & 337.99   & 0.1852  & $0.40\pm0.15$    \\
     &          & NIRC2 & $K^\prime$ & B & $336\pm3$ & $0.175\pm0.015$ & $0.11\pm0.04$ \\
     &          & UKIRT & $J$       & C & 45.0     & 3.05    & $4.761\pm0.063$   \\

3284 &  6497146 & DSSI  & 692nm     & B & 193.06   & 0.4380  & $3.56\pm0.15$    \\
     &          & NIRC2 & $K^\prime$ & B & \nodata  & \nodata & $2.01\pm0.15$    \\
     &          & WIYN  & $B$       & C & \nodata  & \nodata & $1.802\pm0.046$   \\
     &          & WIYN  & $V$       & C & 3.2      & 3.98    & $2.013\pm0.035$   \\
     &          & UKIRT & $J$       & C & 3.2      & 4.01    & $2.904\pm0.008$   \\

4407 &  8396660 & Pharo & $Ks$      & B & 299.8    & 2.45    & $1.988\pm0.005$   \\
     &          & Pharo & $Ks$      & C & 311.0    & 2.65    & $4.972\pm0.022$   \\
\enddata
\tablenotetext{a}{DSSI = Differential Speckle Survey Instrument at Gemini North; Pharo = Near-IR AO imager at Palomar 5m;
  NIRC2 = Near-IR AO imager at Keck~II; UKIRT = $J$-band survey at UKIRT (Phil Lucas, from cfop.ipac.caltech.edu); 
  WIYN = Mosaic2.0 Camera at WIYN~0.9m \citep{everettetal12}}
\tablenotetext{b}{Astrometry based on combination of $J$ and $Ks$ filters}
\tablenotetext{c}{Also known as Kepler-132}
\end{deluxetable}

\begin{deluxetable}{ccccccccc}
\tabletypesize{\scriptsize}
\tablecolumns{10}
\tablewidth{0pt}
\tablecaption{KOI Neighbors Considered as Bound Companions\label{table:isofits}}
\tablehead{
  \colhead{KOI} & \colhead{component} & \colhead{$\theta$($\arcdeg$)} & \colhead{$\rho$($\arcsec$)} & \colhead{apparent $Kp$\tablenotemark{a}} &
  \colhead{$P_{BG}$} & \colhead{Filter} & \colhead{$Kp$\tablenotemark{b}} &
 \colhead{ $\frac{Kp - <Kp>}{\sigma (Kp)} ^b$ } \\
}
\startdata
268 & KOI & \nodata & \nodata & 10.56 & \nodata & \nodata & $3.56\pm0.10$ & \nodata \\[10pt]
268 & B & 267.69 & 1.7591 & 14.88 & $5.1\times10^{-4}$ & $Ks$ & $7.87\begin{array}{c}+0.13\\-0.11\end{array}$ & -0.10 \\[10pt]
268 & B & 267.69 & 1.7591 & 14.88 & \nodata & $J$ & $7.89\begin{array}{c}+0.14\\-0.10\end{array}$ & 0.07 \\[10pt]
268 & C & 310.19 & 2.5243 & 16.59 & $2.5\times10^{-3}$ & $Ks$ & $9.59\begin{array}{c}+0.09\\-0.07\end{array}$ & 0.0 \\[10pt]
268 & C & 310.19 & 2.5243 & 16.59 & \nodata & $J$ & $9.59\begin{array}{c}+0.11\\-0.09\end{array}$ & 0.0 \\[10pt]
284\tablenotemark{c} & KOI & \nodata & \nodata & 12.38 & \nodata & \nodata & $3.75\begin{array}{c}+0.41\\-0.40\end{array}$ & \nodata \\[10pt]
284 & B & 97.44 & 0.867 & 12.80 & $4.8\times10^{-5}$ & $Ks$ & $4.02\begin{array}{c}+0.45\\-0.46\end{array}$ & -0.32 \\[10pt]
284 & B & 97.44 & 0.867 & 12.80 & \nodata & $J$ & $4.02\begin{array}{c}+0.42\\-0.44\end{array}$ & -0.34 \\[10pt]
284 & B & 97.44 & 0.867 & 12.80 & \nodata & 692nm & $4.41\begin{array}{c}+0.44\\-0.40\end{array}$ & 0.61 \\[10pt]
1964 & KOI & \nodata & \nodata & 10.73 & \nodata & \nodata & $4.78\begin{array}{c}+0.23\\-0.26\end{array}$ & \nodata \\[10pt]
1964 & B & 2.26 & 0.3978 & 14.10 & $1.2\times10^{-5}$ & $Ks$ & $8.00\begin{array}{c}+0.33\\-0.34\end{array}$ & -0.46 \\[10pt]
1964 & B & 2.26 & 0.3978 & 14.10 & \nodata & $J$ & $7.64\begin{array}{c}+0.30\\-0.36\end{array}$ & -1.74 \\[10pt]
1964 & B & 2.26 & 0.3978 & 14.10 & \nodata & 880nm & $8.36\begin{array}{c}+0.30\\-0.29\end{array}$ & 0.70 \\[10pt]
1964 & B & 2.26 & 0.3978 & 14.10 & \nodata & 692nm & $8.38\begin{array}{c}+0.23\\-0.22\end{array}$ & 1.05 \\[10pt]
2311 & KOI & \nodata & \nodata & 12.57 & \nodata & \nodata & $4.29\begin{array}{c}+0.39\\-0.37\end{array}$ & \nodata \\[10pt]
2311 & B & 70.19 & 1.0264 & 19.02 & $2.1\times10^{-3}$ & $Ks$ & $11.32\begin{array}{c}+0.37\\-0.46\end{array}$ & 1.26 \\[10pt]
2311 & B & 70.19 & 1.0264 & 19.02 & \nodata & $J$ & $11.59\begin{array}{c}+0.40\\-0.44\end{array}$ & 1.92 \\[10pt]
2311 & B & 70.19 & 1.0264 & 19.02 & \nodata & 692nm & $9.72\begin{array}{c}+0.38\\-0.36\end{array}$ & -2.70 \\[10pt]
3255 & KOI & \nodata & \nodata & 14.92 & \nodata & \nodata & $7.04\begin{array}{c}+0.22\\-0.23\end{array}$ & \nodata \\[10pt]
3255 & B & 337.20 & 0.1832 & 15.33 & $1.3\times10^{-5}$ & $Ks$ & $7.25\pm0.21$ & -0.90 \\[10pt]
3255 & B & 337.20 & 0.1832 & 15.33 & \nodata & 880nm & $7.54\begin{array}{c}+0.27\\-0.22\end{array}$ & 0.46 \\[10pt]
3255 & B & 337.20 & 0.1832 & 15.33 & \nodata & 692nm & $7.54\begin{array}{c}+0.26\\-0.22\end{array}$ & 0.46 \\[10pt]
3255 & C & 45.0 & 3.05 & 19.77 & $6.2\times10^{-2}$ & $J$ & $12.73\begin{array}{c}+0.24\\-0.13\end{array}$ & \nodata \\[10pt]
3284 & KOI & \nodata & \nodata & 14.55 & \nodata & \nodata & $8.89\begin{array}{c}+0.17\\-0.22\end{array}$ & \nodata \\[10pt]
3284 & B & 193.06 & 0.4380 & 17.32 & $4.6\times10^{-5}$ & $Ks$ & $11.28\pm0.19$ & -2.07 \\[10pt]
3284 & B & 193.06 & 0.4380 & 17.32 & \nodata & 692nm & $12.29\begin{array}{c}+0.19\\-0.24\end{array}$ & 2.62 \\[10pt]
3284 & C & 3.2 & 4.01 & 16.73 & $9.2\times10^{-3}$ & $J$ & $12.35\begin{array}{c}+0.13\\-0.18\end{array}$ & 6.53 \\[10pt]
3284 & C & 3.2 & 4.01 & 16.73 & \nodata & $V$ & $10.71\begin{array}{c}+0.19\\-0.24\end{array}$ & -2.33 \\[10pt]
3284 & C & 3.2 & 4.01 & 16.73 & \nodata & $B$ & $10.45\begin{array}{c}+0.17\\-0.24\end{array}$ & -4.02 \\[10pt]
4407 & KOI & \nodata & \nodata & 11.18 & \nodata & \nodata & $3.34\begin{array}{c}+0.44\\-0.45\end{array}$ & \nodata \\[10pt]
4407 & B & 299.8 & 2.45 & 14.36 & $1.9\times10^{-3}$ & $Ks$ & $6.52\begin{array}{c}+0.89\\-1.01\end{array}$ & \nodata \\[10pt]
4407 & C & 311.0 & 2.65 & 18.64 & $2.5\times10^{-2}$ & $Ks$ & $10.80\begin{array}{c}+0.49\\-0.54\end{array}$ & \nodata \\[10pt]

\enddata
\tablenotetext{a}{In this column $Kp$ magnitudes for secondary stars are
mean values based on a combination of all filters (\S\ref{subsubsec:secondaryprops}).}
\tablenotetext{b}{In these columns $K_p$ refers to absolute \kepler magnitudes, which are
calculated independently for each filter.}
\tablenotetext{c}{Also known as Kepler-132}
\end{deluxetable}

\begin{deluxetable}{llccccccc}
\tabletypesize{\scriptsize}
\tablecolumns{10}
\tablewidth{0pt}
\tablecaption{Magnitudes and Colors of KOI Stars and their Neighbors\tablenotemark{a}\label{table:neighborcolors}}
\tablehead{
  \colhead{KOI} & \colhead{component} & \colhead{$692nm$} & \colhead{$880nm$} &
  \colhead{$J$} & \colhead{$Ks$} & \colhead{$692nm-880nm$} & \colhead{$692nm-Ks$} & \colhead{$J-Ks$} \
}
\startdata
268 & KOI & $3.58\begin{array}{c}+0.10\\-0.09\end{array}$ & $3.63\begin{array}{c}+0.10\\-0.08\end{array}$ & $2.82\begin{array}{c}+0.09\\-0.07\end{array}$ & $2.55\begin{array}{c}+0.09\\-0.06\end{array}$ & $0.05\begin{array}{c}+0.00\\-0.01\end{array}$ & $1.03\begin{array}{c}+0.02\\-0.04\end{array}$ & $0.27\pm0.01$ \\[10pt]
268 & B & \nodata & \nodata & $5.93\begin{array}{c}+0.10\\-0.09\end{array}$ & $5.09\begin{array}{c}+0.09\\-0.07\end{array}$ & \nodata & \nodata & $0.84\pm0.06$ \\[10pt]
268 & C & \nodata & \nodata & $7.15\begin{array}{c}+0.10\\-0.09\end{array}$ & $6.34\begin{array}{c}+0.10\\-0.07\end{array}$ & \nodata & \nodata & $0.81\begin{array}{c}+0.06\\-0.07\end{array}$ \\[10pt]
284\tablenotemark{b} & KOI & $3.73\begin{array}{c}+0.41\\-0.39\end{array}$ & $3.73\begin{array}{c}+0.41\\-0.40\end{array}$ & $2.87\begin{array}{c}+0.41\\-0.40\end{array}$ & $2.53\begin{array}{c}+0.41\\-0.40\end{array}$ & $0.01\pm0.01$ & $1.20\pm0.03$ & $0.34\pm0.01$ \\[10pt]
284 & B & $4.39\begin{array}{c}+0.44\\-0.42\end{array}$ & \nodata & $3.13\begin{array}{c}+0.41\\-0.40\end{array}$ & $2.79\begin{array}{c}+0.41\\-0.40\end{array}$ & \nodata & $1.60\pm0.16$ & $0.34\pm0.04$ \\[10pt]
1964 & KOI & $4.73\begin{array}{c}+0.23\\-0.26\end{array}$ & $4.69\begin{array}{c}+0.23\\-0.25\end{array}$ & $3.77\begin{array}{c}+0.23\\-0.24\end{array}$ & $3.37\pm0.23$ & $0.04\pm0.01$ & $1.36\begin{array}{c}+0.04\\-0.05\end{array}$ & $0.41\pm0.02$ \\[10pt]
1964 & B & $8.27\begin{array}{c}+0.27\\-0.30\end{array}$ & $7.54\begin{array}{c}+0.27\\-0.29\end{array}$ & $5.73\begin{array}{c}+0.23\\-0.24\end{array}$ & $5.15\pm0.23$ & $0.73\pm0.21$ & $3.12\pm0.16$ & $0.59\pm0.04$ \\[10pt]
2311 & KOI & $4.24\begin{array}{c}+0.39\\-0.37\end{array}$ & $4.23\begin{array}{c}+0.39\\-0.37\end{array}$ & $3.33\begin{array}{c}+0.39\\-0.37\end{array}$ & $2.95\begin{array}{c}+0.39\\-0.37\end{array}$ & $0.02\pm0.01$ & $1.29\pm0.04$ & $0.37\pm0.02$ \\[10pt]
2311 & B & $9.71\begin{array}{c}+0.41\\-0.40\end{array}$ & \nodata & $8.71\begin{array}{c}+0.41\\-0.39\end{array}$ & $7.69\begin{array}{c}+0.39\\-0.37\end{array}$ & \nodata & $2.02\pm0.17$ & $1.01\pm0.14$ \\[10pt]
3255 & KOI & $6.85\begin{array}{c}+0.22\\-0.21\end{array}$ & $6.61\begin{array}{c}+0.17\\-0.18\end{array}$ & $5.43\begin{array}{c}+0.16\\-0.17\end{array}$ & $4.69\begin{array}{c}+0.13\\-0.15\end{array}$ & $0.24\begin{array}{c}+0.06\\-0.04\end{array}$ & $2.16\begin{array}{c}+0.10\\-0.11\end{array}$ & $0.75\begin{array}{c}+0.03\\-0.04\end{array}$ \\[10pt]
3255 & B & $7.37\pm0.26$ & $7.01\begin{array}{c}+0.23\\-0.24\end{array}$ & \nodata & $4.80\begin{array}{c}+0.14\\-0.16\end{array}$ & $0.36\pm0.22$ & $2.57\begin{array}{c}+0.18\\-0.19\end{array}$ & \nodata \\[10pt]
3284 & KOI & $8.79\begin{array}{c}+0.18\\-0.24\end{array}$ & $7.99\begin{array}{c}+0.14\\-0.18\end{array}$ & $6.64\begin{array}{c}+0.13\\-0.16\end{array}$ & $5.83\begin{array}{c}+0.13\\-0.17\end{array}$ & $0.80\begin{array}{c}+0.06\\-0.07\end{array}$ & $2.97\begin{array}{c}+0.07\\-0.09\end{array}$ & $0.82\pm0.01$ \\[10pt]
3284 & B & $12.35\begin{array}{c}+0.23\\-0.28\end{array}$ & \nodata & \nodata & $7.84\begin{array}{c}+0.14\\-0.17\end{array}$ & \nodata & $4.52\begin{array}{c}+0.17\\-0.18\end{array}$ & \nodata \\[10pt]
\enddata
\tablenotetext{a}{For KOI stars (labelled as component KOI), magnitudes are absolute values from isochrone fits.  For neighbors (labelled as component B or C), magnitudes represent absolute values only if they lie at the same distance as the KOI star.}
\tablenotetext{b}{Also known as Kepler-132}
\end{deluxetable}

\begin{deluxetable}{ccccccc}
\tabletypesize{\scriptsize}
\tablecolumns{10}
\tablewidth{0pt}
\tablecaption{Blended Candidate Host Stars (Bound Companion Case)\label{table:blendedstars}}
\tablehead{
  \colhead{KOI} & \colhead{star} & \colhead{$\theta$} & \colhead{$\rho$} & 
  \colhead{$\Delta Kp$} & \colhead{flux in aperture} & \colhead{$\delta_{true}/\delta_{obs}$} \\
  \colhead{} & \colhead{} & \colhead{($\arcdeg$)} & \colhead{($\arcsec$)} & 
  \colhead{} & \colhead{(\%)} & \colhead{} \\
}
\startdata

268.01  & KOI 268     & \nodata & \nodata & 0.00   & 98.358 & 1.022  \\      
268.01  & B           & 267.69  & 1.7591  & 4.32   & 98.331 & 55.28  \\       
268.01  & C           & 310.19  & 2.5243  & 6.03   & 98.076 & 264.9  \\       
                                                                                       
284.01  & Kepler-132  & \nodata & \nodata & 0.00   & 95.270 & 1.685  \\        
284.01  & B           & 97.44   & 0.867   & 0.41   & 95.148 & 2.461  \\        
284.02  & Kepler-132  & \nodata & \nodata & 0.00   & 95.336 & 1.685  \\        
284.02  & B           & 97.44   & 0.867   & 0.41   & 95.206 & 2.461  \\        
284.03  & Kepler-132  & \nodata & \nodata & 0.00   & 95.364 & 1.684  \\        
284.03  & B           & 97.44   & 0.867   & 0.41   & 95.183 & 2.462  \\        
284.04  & Kepler-132  & \nodata & \nodata & 0.00   & 95.193 & 1.684  \\        
284.04  & B           & 97.44   & 0.867   & 0.41   & 94.974 & 2.462  \\        
                                                                                       
1964.01 & KOI 1964    & \nodata & \nodata & 0.00   & 98.202 & 1.045  \\        
1964.01 & B           & 2.26    & 0.3978  & 3.37   & 98.070 & 23.30  \\        
                                                                                       
2311.01 & KOI 2311    & \nodata & \nodata & 0.00   & 94.821 & 1.003  \\     
2311.01 & B           & 70.19   & 1.0264  & 6.45   & 94.149 & 384.0  \\     
2311.02 & KOI 2311    & \nodata & \nodata & 0.00   & 94.741 & 1.003  \\     
2311.02 & B           & 70.19   & 1.0264  & 6.45   & 94.275 & 383.1  \\     
                                                                                       
3255.01 & KOI 3255    & \nodata & \nodata & 0.00   & 87.770 & 1.700  \\        
3255.01 & B           & 337.20  & 0.1832  & 0.41   & 87.768 & 2.465  \\        
3255.01 & C           & 45.0    & 3.05    & 5.69   & 69.447 & 189.0  \\        
                                                                                       
3284.01 & KOI 3284    & \nodata & \nodata & 0.00   & 68.748 & 1.160  \\    
3284.01 & B           & 193.06  & 0.4380  & 2.77   & 68.832 & 14.86  \\    
3284.01 & C           & 3.2     & 4.01    & 2.26   & 45.582 & 16.02  \\    
                                                                                       
4407.01 & KOI 4407    & \nodata & \nodata & 0.00   & 96.491 & 1.054  \\     
4407.01 & B           & 299.8   & 2.45    & 3.18   & 96.426 & 19.74  \\    
4407.01 & C           & 311.0   & 2.65    & 7.46   & 96.034 & 1021   \\     
\enddata
\end{deluxetable}

\begin{deluxetable}{cccccccc}
\tabletypesize{\scriptsize}
\tablecolumns{10}
\tablewidth{0pt}
\tablecaption{Blended Candidate Host Stars (Field Star Cases)\label{table:blendedBGstars}}
\tablehead{
  \colhead{KOI} & \colhead{star} & \colhead{status\tablenotemark{a}} & \colhead{$\theta$} & \colhead{$\rho$} & 
  \colhead{$\Delta Kp$} & \colhead{flux in aperture} & \colhead{$\delta_{true}/\delta_{obs}$} \\
  \colhead{} & \colhead{} & \colhead{} & \colhead{($\arcdeg$)} & \colhead{($\arcsec$)} & 
  \colhead{} & \colhead{(\%)} & \colhead{} \\
}
\startdata

1964.01 & KOI 1964 & \nodata      & \nodata & \nodata & 0.00   & 98.202 & 1.106  \\        
1964.01 & B        & background   & 2.26    & 0.3978  & 2.44   & 98.070 & 10.47  \\        
                                                                                       
2311.01 & KOI 2311 & \nodata      & \nodata & \nodata & 0.00   & 94.821 & 1.005  \\     
2311.01 & B        & background   & 70.19   & 1.0264  & 5.70   & 94.149 & 192.9  \\     
2311.02 & KOI 2311 & \nodata      & \nodata & \nodata & 0.00   & 94.741 & 1.005  \\     
2311.02 & B        & background   & 70.19   & 1.0264  & 5.70   & 94.275 & 192.5  \\     

3255.01 & KOI 3255 & \nodata      & \nodata & \nodata & 0.00   & 87.770 & 1.698  \\
3255.01 & B        & bound        & 337.20  & 0.1832  & 0.41   & 87.768 & 2.466  \\                
3255.01 & C        & background   & 45.0    & 3.05    & 4.81   & 69.447 & 182.7  \\        
                                                                                       
3284.01 & KOI 3284 & \nodata      & \nodata & \nodata & 0.00   & 68.748 & 1.145  \\    
3284.01 & B        & bound        & 193.06  & 0.4380  & 2.77   & 68.832 & 14.66  \\    
3284.01 & C        & background   & 3.2     & 4.01    & 2.48   & 45.582 & 19.42  \\    

4407.01 & KOI 4407 & \nodata      & \nodata & \nodata & 0.00   & 96.491 & 1.084  \\     
4407.01 & B        & background   & 299.8   & 2.45    & 2.70   & 96.426 & 13.04  \\    
4407.01 & C        & background   & 311.0   & 2.65    & 7.64   & 96.034 & 1239   \\     
                                                                 
\enddata
\tablenotetext{a}{Status indicates whether neighboring stars are assumed
  to be background stars or gravitationally-bound companions for the
  purpose of determining relative \kepler magnitudes in this table.}
\end{deluxetable}

\begin{deluxetable}{ccccccccccccc}
\tablecolumns{13}
\tabletypesize{\scriptsize}
\tablewidth{0pt}
\tablecaption{Planet Properties for KOIs and Neighbor Stars as Hosts\label{table:planetproperties}}
\tablehead{
  \colhead{KOI} & \colhead{component} & \colhead{planet} & \colhead{Kepler Name} &
 \colhead{$R_\star$\tablenotemark{a}} & \colhead{$M_\star$\tablenotemark{a}} & \colhead{$T_{\rm eff}$\tablenotemark{a}} &
 \colhead{$P_p$} & \colhead{$R_p/R_\star$} &
  \colhead{$a$\tablenotemark{b}} & \colhead{$T_{eq}$\tablenotemark{b}} & \colhead{$R_{p{\rm BIN}}$\tablenotemark{b,c}} & \colhead{$R_{p{\rm BG}}$\tablenotemark{b,c}} \\
 \colhead{} & \colhead{} & \colhead{} & \colhead{} &
 \colhead{($R_\odot$)} & \colhead{($M_\odot$)} & \colhead{(K)} &
 \colhead{(days)} & &
  \colhead{(AU)} & \colhead{(K)} & \colhead{($R_\oplus$)} & \colhead{($R_\oplus$)} \\
}
\startdata
 115 & KOI &  115.01 & Kepler-105b & 1.015 & 1.027 & 6065 &   5.412 & $2.292\times10^{-2}$ & 0.0609 & 1092 & 2.54 & \nodata \\ 
 115 & KOI &  115.02 & Kepler-105c & 1.015 & 1.027 & 6065 &   7.126 & $1.296\times10^{-2}$ & 0.0731 &  997 & 1.44 & \nodata \\ 
 265 & KOI &  265.01 & \nodata & 1.564 & 1.097 & 5915 &   3.568 & $1.001\times10^{-2}$ & 0.0471 & 1503 & 1.71 & \nodata \\ 
 268 & KOI &  268.01 & \nodata & 1.366 & 1.230 & 6343 & 110.379 & $2.011\times10^{-2}$ & 0.4825 &  470 & 3.04 & \nodata \\ 
 268 &   B &  268.01 & \nodata & 0.571 & 0.596 & 4007 & 110.379 & $2.011\times10^{-2}$ & 0.3789 &  217 & 9.33 & \nodata \\ 
 274 & KOI &  274.01 & Kepler-128b & 1.659 & 1.184 & 6072 &  15.092 & $6.630\times10^{-3}$ & 0.1264 &  970 & 1.20 & \nodata \\ 
 274 & KOI &  274.02 & Kepler-128c & 1.659 & 1.184 & 6072 &  22.795 & $6.670\times10^{-3}$ & 0.1664 &  845 & 1.21 & \nodata \\ 
 284 & KOI &  284.01 & Kepler-132d & 1.408 & 1.023 & 5879 &  18.010 & $1.490\times10^{-2}$ & 0.1355 &  836 & 2.98 & \nodata \\ 
 284 &   B &  284.01 & Kepler-132d & 1.169 & 0.987 & 5850 &  18.010 & $1.490\times10^{-2}$ & 0.1339 &  762 & 2.99 & \nodata \\ 
 284 & KOI &  284.02 & Kepler-132c & 1.408 & 1.023 & 5879 &   6.415 & $1.135\times10^{-2}$ & 0.0681 & 1179 & 2.27 & \nodata \\ 
 284 &   B &  284.02 & Kepler-132c & 1.169 & 0.987 & 5850 &   6.415 & $1.135\times10^{-2}$ & 0.0673 & 1075 & 2.27 & \nodata \\ 
 284 & KOI &  284.03 & Kepler-132b & 1.408 & 1.023 & 5879 &   6.178 & $1.087\times10^{-2}$ & 0.0664 & 1194 & 2.17 & \nodata \\ 
 284 &   B &  284.03 & Kepler-132b & 1.169 & 0.987 & 5850 &   6.178 & $1.087\times10^{-2}$ & 0.0656 & 1089 & 2.18 & \nodata \\ 
 284 & KOI &  284.04 & \nodata & 1.408 & 1.023 & 5879 & 110.287 & $1.125\times10^{-2}$ & 0.4535 &  457 & 2.25 & \nodata \\ 
 284 &   B &  284.04 & \nodata & 1.169 & 0.987 & 5850 & 110.287 & $1.125\times10^{-2}$ & 0.4481 &  416 & 2.25 & \nodata \\ 
 369 & KOI &  369.01 & Kepler-144b & 1.491 & 1.126 & 6157 &   5.885 & $1.090\times10^{-2}$ & 0.0664 & 1287 & 1.78 & \nodata \\ 
 369 & KOI &  369.02 & Kepler-144c & 1.491 & 1.126 & 6157 &  10.105 & $1.037\times10^{-2}$ & 0.0952 & 1075 & 1.69 & \nodata \\ 
1537 & KOI & 1537.01 & \nodata & 1.824 & 1.366 & 6260 &  10.191 & $6.750\times10^{-3}$ & 0.1021 & 1167 & 1.35 & \nodata \\ 
1964 & KOI & 1964.01 & \nodata & 0.989 & 0.871 & 5547 &   2.229 & $6.910\times10^{-3}$ & 0.0319 & 1362 & 0.764 & 0.785 \\ 
1964 &   B & 1964.01 & \nodata & 0.556 & 0.569 & 3892 &   2.229 & $6.910\times10^{-3}$ & 0.0277 &  769 & 2.03 & \nodata \\ 
2311 & KOI & 2311.01 & \nodata & 1.182 & 0.975 & 5657 & 191.864 & $8.900\times10^{-3}$ & 0.6456 &  337 & 1.15 & 1.15 \\ 
2311 &   B & 2311.01 & \nodata & 0.270 & 0.256 & 3285 & 191.864 & $8.900\times10^{-3}$ & 0.4135 &  117 & 5.14 & \nodata \\ 
2311 & KOI & 2311.02 & \nodata & 1.182 & 0.975 & 5657 &  13.726 & $7.200\times10^{-3}$ & 0.1112 &  813 & 0.932 & 0.932 \\ 
2311 &   B & 2311.02 & \nodata & 0.270 & 0.256 & 3285 &  13.726 & $7.200\times10^{-3}$ & 0.0713 &  282 & 4.16 & \nodata \\ 
2365 & KOI & 2365.01 & Kepler-430b & 1.485 & 1.166 & 5884 &  35.968 & $2.003\times10^{-2}$ & 0.2244 &  667 & 3.25 & \nodata \\ 
2365 & KOI & 2365.02 & Kepler-430c & 1.485 & 1.166 & 5884 & 110.979 & $1.080\times10^{-2}$ & 0.4757 &  458 & 1.75 & \nodata \\ 
2593 & KOI & 2593.01 & \nodata & 1.453 & 1.230 & 6119 &  14.798 & $6.910\times10^{-3}$ & 0.1264 &  915 & 1.10 & \nodata \\ 
2755 & KOI & 2755.01 & \nodata & 1.172 & 0.973 & 5792 &   8.483 & $8.300\times10^{-3}$ & 0.0807 &  974 & 1.06 & \nodata \\ 
3097 & KOI & 3097.01 & Kepler-431d & 1.092 & 1.071 & 6004 &  11.922 & $9.300\times10^{-3}$ & 0.1045 &  856 & 1.11 & \nodata \\ 
3097 & KOI & 3097.02 & Kepler-431b & 1.092 & 1.071 & 6004 &   6.803 & $6.400\times10^{-3}$ & 0.0719 & 1032 & 0.764 & \nodata \\ 
3097 & KOI & 3097.03 & Kepler-431c & 1.092 & 1.071 & 6004 &   8.703 & $5.600\times10^{-3}$ & 0.0847 &  951 & 0.668 & \nodata \\ 
3204 & KOI & 3204.01 & \nodata & 1.593 & 1.553 & 7338 &   0.573 & $5.800\times10^{-3}$ & 0.0156 & 3268 & 1.01 & \nodata \\ 
3224 & KOI & 3224.01 & \nodata & 0.962 & 0.866 & 5382 &   3.439 & $6.340\times10^{-3}$ & 0.0425 & 1129 & 0.667 & \nodata \\ 
3255 & KOI & 3255.01 & \nodata & 0.622 & 0.615 & 4427 &  66.651 & $2.385\times10^{-2}$ & 0.2736 &  294 & 2.11 & 2.11 \\ 
3255 &   B & 3255.01 & \nodata & 0.592 & 0.593 & 4227 &  66.651 & $2.385\times10^{-2}$ & 0.2703 &  276 & 2.42 & \nodata \\ 
3284 & KOI & 3284.01 & \nodata & 0.463 & 0.479 & 3688 &  35.233 & $1.830\times10^{-2}$ & 0.1646 &  272 & 0.997 & 0.991 \\ 
3284 &   B & 3284.01 & \nodata & 0.189 & 0.163 & 3255 &  35.233 & $1.830\times10^{-2}$ & 0.1148 &  184 & 1.46 & \nodata \\ 
4407 & KOI & 4407.01 & \nodata & 1.435 & 1.234 & 6408 &   1.338 & $4.040\times10^{-3}$ & 0.0255 & 2121 & 0.650 & 0.660 \\ 
\enddata
\tablenotetext{a}{Stellar radii, masses and effective temperatures, when listed for neighbor stars (labelled component B) assume the neighbor is a bound companion.}
\tablenotetext{b}{Revised planetary properties}
\tablenotetext{c}{Two values of the planet radius are
calculated for KOIs subjected to blending by neighboring stars.
$R_{p{\rm BIN}}$ represents the radius when the neighbors are assumed to
be binary companions.  $R_{p{\rm BG}}$ represents the radius when the
neighbors are assumed to be background stars.}
\end{deluxetable}

\end{document}